\documentclass[a4paper,usenatbib]{mnras}
\usepackage{newtxtext,newtxmath}
\usepackage[T1]{fontenc}
\usepackage{ae,aecompl}
\usepackage{graphicx}	
\usepackage{times}
\usepackage{mathtools}
\usepackage{epsf}
\usepackage{enumitem}
\usepackage{footnote}
\usepackage{relsize}
\usepackage{xfrac}
\usepackage{aasmacros}
\usepackage{ulem}
\usepackage{color}
\usepackage{pgfplots}

\usepackage{tabularx}
\usepackage{setspace}
\usepackage[squaren, Gray, cdot]{SIunits}
\DeclareMathOperator{\sech}{sech}
\definecolor{Pink}{rgb}{1.0,0.05,0.5}

\definecolor{Red}{rgb}{1.0,0.0,0.0}
\definecolor{Green}{rgb}{0.0,1.0,0.0}
\definecolor{Orange}{rgb}{1.0,0.05,0.15}
\definecolor{lblue}{rgb}{0.1,0.7,1.}
\definecolor{lbluel}{rgb}{0.0,0.5,1}
\usepackage{hyperref} 
\hypersetup{colorlinks=true, linkcolor=blue, citecolor=lbluel, filecolor=blue, urlcolor=lbluel}

\def\corrections#1{\noindent{\red [CORRECTIONS #1]}}
\def\thesis#1{\noindent{\green [THESIS CORRECTIONS #1]}}
\def\corrections#1{\noindent{#1}}
\def\thesis#1{\noindent{#1}}

 \newcommand\rono{-2.72}                               
 \newcommand\rTno{-3.36}                               
 \newcommand\rwno{-2.85}                               
 \newcommand\rofb{-2.24}

 \newcommand\ronoDM{-3.18}                             
 \newcommand\rTnoDM{-2.16}                             
 \newcommand\rwnoDM{-1.99}

 \newcommand\ronoerr{0.26}                             
 \newcommand\rTnoerr{0.12}                             
 \newcommand\rwnoerr{0.17}                             
 \newcommand\rofberr{0.34}

 \newcommand\ronoDMerr{0.28}                           
 \newcommand\rTnoDMerr{0.12}                           
 \newcommand\rwnoDMerr{0.09}

 \newcommand\cono{-3.10}                               
 \newcommand\coDMno{-2.07}                             
 \newcommand\conoerr{0.10}                             
 \newcommand\coDMnoerr{0.07}

\begin{document}

\title[High Redshift Filaments]{Rivers of Gas I.: Unveiling The Properties of High Redshift Filaments}
\author[Marius Rams\o y et al.]{Marius Rams\o y$^{1}$\thanks{E-mail: marius.ramsoy@physics.ox.ac.uk}, Adrianne Slyz$^{1}$,  Julien Devriendt$^{1}$, Clotilde Laigle$^{1,2}$ and \newauthor Yohan Dubois$^{2}$
\\
$^{1}$ Sub-department of Astrophysics, University of Oxford, Keble Road, Oxford OX1 3RH, UK \\
$^{2}$ Sorbonne Universit{\'e}s, CNRS, UMR 7095,  Institut d'Astrophysique de Paris, 98 bis bd Arago, 75014 Paris, France 
}
\date{Accepted XXX. Received YYY; in original form ZZZ}
\maketitle
\pubyear{2019}


\label{firstpage}
\pagerange{\pageref{firstpage}--\pageref{lastpage}}

\begin{abstract}
At high redshift, the cosmic web is widely expected to have a significant impact on the morphologies, dynamics and star formation rates of the galaxies embedded within it, underscoring the need for a comprehensive study of the properties of such a filamentary network. 
With this goal in mind, we perform an analysis of high-$z$ gas and dark matter (DM) filaments around a Milky Way-like progenitor simulated with the {\sc ramses} adaptive mesh refinement (AMR) code  from cosmic scales ($\sim$1 Mpc) down to the virial radius of its DM halo host ($\sim$20 kpc at $z=4$).  
Radial density profiles of both gas and DM filaments are found to have the same functional form, namely a plummer-like profile modified to take into account the wall within which these filaments are embedded. 
Measurements of the typical filament core radius $r_0$ from the simulation are consistent with that of isothermal cylinders in hydrostatic equilibrium. 
Such an analytic model also predicts a redshift evolution for the core radius of filaments in fair agreement with the measured value for DM ($r_0 \propto (1+z)^{\ronoDM\pm\ronoDMerr}$). Gas filament cores grow as ($r_0 \propto (1+z)^{\rono\pm\ronoerr}$).
In both gas and DM, temperature and vorticity sharply drop at the edge of filaments, providing an excellent way to constrain the outer filament radius. When feedback is included the gas temperature and vorticity fields are strongly perturbed, hindering such a measurement in the vicinity of the galaxy. However, the core radius of the filaments as measured from the gas density field is largely unaffected by feedback, and the median central density is only reduced by about 20\%.
\end{abstract}




\section{Introduction}\label{Introduction}
Galactic surveys have revealed the presence of anisotropic structure on scales of Mpc, made up of nodes, voids, sheets and filaments \citep[e.g.][]{Davis1982,Lapperent1986,Geller1989}.  Cosmological simulations are able to reproduce this network, the so-called cosmic web \citep{Bond96,Dmitry}, and unveil its existence not just for the distribution of galaxies but also for the underlying gas and DM density,  as a consequence of the hierarchical growth of structures in $\Lambda$CDM.
Gravity amplifies small anisotropies, resulting in a near homogenous background collapsing to form sheets which can collapse again along another axis to form filaments.  Halos form at filament intersections where, according to cosmological hydrodynamics simulations, galaxies at high redshift grow in mass and angular momentum primarily by material transported along these filaments \citep{Birnboim2003,Keres05,Ocvirk2008,Pichon11,Danovich2012,Stewart2013}.

While at large scale gas filaments closely follow the structure of their DM counterparts  in the cosmic web, at the scale of halos they can penetrate deep into the virial radius and even connect to galactic disks triggering star formation episodes \citep[see e.g.][]{Katz03, Keres05,Woods14,Stewart2017}.  The erosion of these small-scale gas filaments at lower redshifts is argued to be at least partly responsible for the bimodal distribution in colour, star formation rates and morphology of galaxies \citep{Dekel06} (though quenching of the largest galaxies are dependent on AGN feedback, see e.g. \cite{Croton2006}). Other implicit evidence for the presence of inflows is the presence of low metallicity G-dwarfs in the solar neighbourhood, as established in the seminal work 
of \cite{vandenBergh1962}. As gas depletion timescales are estimated to be on the order of a few Gyrs for local disk galaxies  \citep[e.g.][]{Bigiel2011,Leroy2013,Rahman2012}, replenishment by inflow of pristine gas is required to match the observations. This finding is also supported by observations of extended gas disks around galaxies (co-rotating with the stellar disk), either directly in emission \citep[e.g. from Lyman-$\alpha$,][]{Prescott2015} or indirectly in absorption \cite[e.g. from galaxy-quasar pairs, as studied in][]{Zabl2019,Ho2019}, all suggesting filamentary accretion from the cosmic web.

\begin{figure*}
\begin{tikzpicture}
\begin{scope}[xshift=-1.5cm]
    \node[anchor=south west,inner sep=0] (image) at (0,0) {\includegraphics[width=\textwidth]{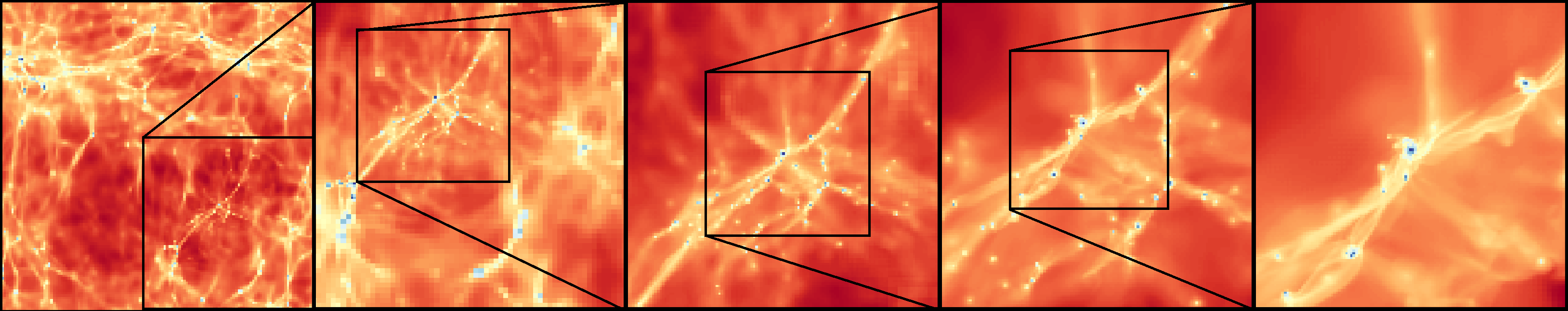}};
   
\end{scope}
\node [anchor=west] (note) at (0.4,1.75) {1.25 Mpc};
\node [anchor=west] (water) at (2.9,2.95) {625 kpc};
\node [anchor=west] (fire) at (6.8,2.5) {313 kpc};
\node [anchor=west] (earth) at (10.2,2.75) {156 kpc};

\end{tikzpicture}
\caption{Zooming in on the NUT galaxy gas density field at $z=4$. The leftmost panel shows a gas density projection of the entire simulation volume (12.5 comoving Mpc), with the high resolution zoom region enclosed in the square located in the bottom right corner of the first panel. Each subsequent panel, going from left to right, displays a projection of 1/8$^\mathrm{th}$ of the volume of the previous panel.  The size of each volume in physical units is indicated.  The \textit{middle} panel shows the region within which the analysis in this paper is performed, chosen so as to maximize the length of the studied filament.     }
    \label{fig:zoom}
\end{figure*}
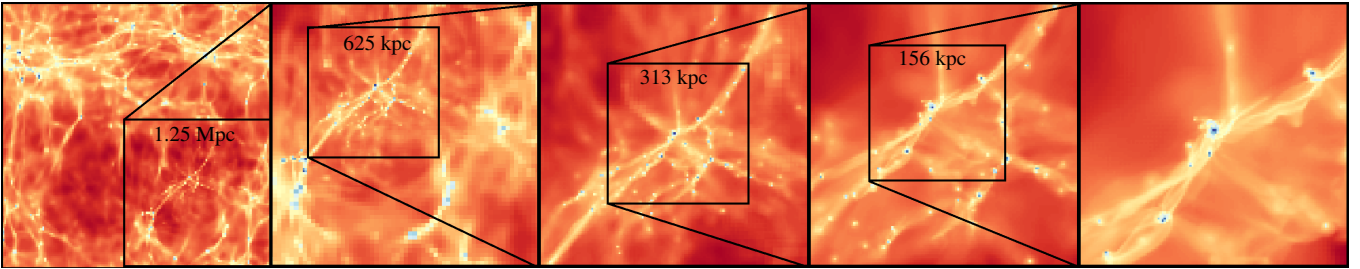
Rather than directly pursuing the filament properties themselves, it is possible to infer them through indirect methods.  On large-scales, many authors have measured halo or galaxy spin alignment with  cosmic filaments both in simulations \citep[see e.g.][]{AragonCalvo2007,Codis12,Dubois2014, Laigle15,Ganesh2018,kraljic19} and low-$z$ spectroscopic observations \citep[e.g.][among others]{Tempel13,Chen2018,Krolewski}. These results highlight a redshift and mass dependence of the alignment signal, with halos with masses above $M_{\rm h}>10^{12} {\rm M}_{\odot}$  displaying spins perpendicularly oriented with respect to the nearest filament, whereas spins of halos with masses below $M_{\rm h}<10^{12} {\rm M}_{\odot}$ align with the nearest filament.  At low masses this is thought to be due to accretion of vorticity rich gas that drive spins to align with the filament.  At high masses this behaviour is overcome by mergers, or as \cite{Laigle15} argues, the accretion of material from multiple vorticity domains.  This dichotomy in galaxy spins shows the profound impact of cosmic filaments on the galaxies embedded within them.  
\corrections{The inverse problem has also been studied \citep{Pandya2019} attempting to use the alignment of galaxies to detect the cosmic web with the CANDELS survey \citep{Grogin2011,Kroekemoer2011}.  The non-detection of the alignment signal is likely due to the number of prolate galaxies with spectroscopically determined redshifts with stellar masses
$9<{\rm log(M_*/M_\odot)}<10$
in the survey, as well as these galaxies' nearest neighbours.
}.

On smaller scales, the misalignment of gas and DM angular momenta in simulations has been attributed to different redistribution processes during halo virialisation \citep[e.g.][]{Kimm2011,Stewart2013}. However, it has also been argued that instabilities within the filaments could develop, leading to their fragmentation and breakup, thereby preventing cold gas from being smoothly accreted by the host galaxy. In such a scenario, the angular momentum segregation between DM and gas could be construed as an artefact of poor numerical resolution in filaments.  
Several authors \citep{Freundlich2014,Mandelker2016,Padnos2018,Mandelker2019,Berlok2019} carried out idealised simulations of filaments entering a halo, and concluded that they should be stable, given their width and velocity. \cite{Cornuault2018} used a phenomenological model of a gas stream to explore the possibility of a turbulent, multi-phase filament.  The accretion efficiency of such a filament would be reduced, but it remains unclear as to whether such a multi-phase model constitutes an acceptable description of cosmological filaments. Using a cosmological zoom simulation tailored to achieve maximum resolution in the filaments, \cite{Rosdahl2012} find that they remain stable within halos with masses of up to a few $10^{11} {\rm M}_{\odot}$ at least as to low as $z=3$, whilst they show more disruption within halos of larger masses \citep[in line with arguments made in][]{Birnboim2003}.  

Ultimately, to distinguish between these scenarios and better assess the role played by filaments on galaxy evolution, quantitative direct measurements of their properties need to be made. However, these have proven notoriously elusive so far \citep[see e.g.][for a more detailed discussion]{Kimm2011}. Indeed, direct observations of the distant cosmic web suffer from the steep scaling of surface brightness with redshift, which makes the cold filaments extremely hard to detect in emission \citep[though not impossible, see e.g.][]{Giavalisco11,Ribaudo11,Kaz12,Martin16,Gallego17,Lydia2020}, and thus rely on stacking, or back-lighting by a bright source. \corrections{Future surveys using the James Webb Space Telescope (JWST) \citep{Gardner2006} while being insensitive to the smoothly accreting gas itself, the telescope will be sensitive to a range of associated phenomena.  Filaments are typically traced by Lyman-$\alpha$ blobs (LABs) and emitters (LAEs) \citep[e.g.][]{Kikuta2019,Umehata2019}, which will be observable by the NIRspec instrument \citep{Latif2011}.  In addition, LAEs should be detectable with the proposed {\sc BlueMUSE} instrument.  On larger scales, filamentary gas can be detected in the radio with the Square Kilometer Array \citep{Kooistra_2019}.  Lyman-$\alpha$ forest tomography also allows the probing of the cosmic web in the IGM, with the feasibility of observations  for the Very Large Telescope investigated by \cite[][]{Lee2014} and the European Extremely Large Telescope by \citep{Japelj2019}.  This will enable the detection and exploration of the full 3 dimensional structure of the cosmic web.}

Efforts to understand 
observed filament properties are correspondingly mirrored by simulations \citep[e.g.][]{Gheller2015,Gheller2016}. On large scales, filaments of the cosmic web are reported to have a radial power law profile in density with a power law index comprised between -1 and -2 \citep[see e.g.][]{Colberg2005,Dolag2006,AragonCalvo2010}.   Smaller scale studies have been performed by e.g. \cite{Ocvirk2016}, who determined the outer radii of filaments in their simulation to be about 50$h^{-1}$kpc at $z = 4.3$ by looking at the separation between temperature peaks caused by the accretion shock, although these authors acknowledge that they did not separate edge-on sheets from filaments in their sample. Using cosmological simulations, \cite{Dekel2009} found that DM filament radii are comparable to the virial radius of the halos they connect, and that the cold gas streams residing within the halos are considerably narrower, typically a few percent of the virial radius.  

\corrections{The scale free nature of CDM results in progressively smaller filaments feeding into larger ones at all scales, down to the numerical resolution of the simulation in this case.  However, alternative DM theories could result in different DM structures, \cite[e.g. Warm DM][]{Gao2007}.  This produces filaments down to Mpc-scales, while erasing smaller scale structure.  \cite{Mocz2019} studied filaments under the Fuzzy DM  regime, where an additional quantum pressure prevents the formation of lower mass filaments.  Both WDM and FDM result in higher density filaments at earlier times, with the formation of population III occurring within them \citep{Mocz2019}, further distinguishing these versions of DM from CDM.  It is possible that the supernovae of these stars will be detectable with JWST \citep{Hartwig_2018}.}

To date, the rich complexity of the filamentary network connecting halos of various masses and its evolution with redshift has yet to be investigated systematically. In this paper we argue that to do so, it is pivotal to work on a cosmological sample of \textit{well-resolved} filaments and take a step in this direction by measuring filament profiles from the density, vorticity, and temperature field information available in a zoom-in cosmological simulation. Our focus is on intermediate-scale filaments, that is, those connecting to a M$_{\star}$ galaxy, at moderate to high redshift ($z \geq 3$). We also investigate how stellar feedback can perturb 
these filaments. Given the limited sample considered in this work, it should be considered a pilot study. In an upcoming paper, the methods developed here will be applied to New-Horizon, a larger volume simulation ($\sim 4200$ (Mpc/h)$^{3}$) with similar resolution \citep[][Dubois et al. in prep]{park2019}, where a statistical sample of filaments can be obtained, connecting a more diverse ensemble of galaxies.  

The structure of this paper is as follows: in section \ref{sec:Simulations} we outline the simulation set up.  In section \ref{sec:fil} we describe how we identify the filaments and perform the analysis. Section \ref{sec:Results}  presents the results of our work, compares filament properties to an analytic model and discusses the robustness of our measurements vis-\`a-vis resolution. We summarize our results in section \ref{sec:Conclusions}.

\section{Numerical methods}
\label{sec:Methods}
\subsection{Simulations}
\label{sec:Simulations}
The analysis is performed on two simulations of the {\sc nut} suite \citep{Powell11}, a series of  cosmological zoom-in simulations of a Milky Way like galaxy designed to study the effects of resolution and various physical processes on its formation and evolution using the Adaptive Mesh Refinement (AMR) code {\sc ramses} \citep{Teyssier02}.  Initial conditions are generated at redshift $z=499$ using the MPGrafic code \citep{Prunet2008} with cosmological parameters set in accordance with the WMAP5 results \citep{Dunkley09}. The simulation volume is a cubic box 9$h^{-1}$Mpc on a side and a coarse root grid of 128$^3$ cells.  A series of three nested grids are then centred on a sphere with radius 2.7$h^{-1}$Mpc which encompasses the Lagrangian volume occupied by the galaxy (host dark matter halo mass of $M_{\rm vir}=5 \times 10^{11} {\rm M}_{\odot}$ by $z=0$). AMR refinement is then enabled within that sphere using a quasi-Lagrangian refinement criterion to achieve a maximal spatial resolution of 10pc at all times whilst forcing the mass of each individual cell to remain roughly constant. The collisionless fluid in this high resolution region consists of dark matter (DM) particles each with mass $5.6 \times 10^4 M_{\odot}$, whereas the gas evolution equations are solved on the AMR grid by means of a Godunov method (HLLC Riemann solver) with a MinMod limiter to reconstruct variables at cell interfaces.  The gas density field in the simulation is shown in Fig.~\ref{fig:zoom} at $z=4$, gradually zooming in from the full box onto the central galaxy itself. 

In this paper, we use a {\sc nut} simulation with no feedback, and one with mechanical supernova feedback as defined in \cite{KimmNUT2015}. In the following, we refer to these two simulations as the ``no-feedback" and ``feedback" runs respectively.  The feedback recipe of \cite{KimmNUT2015} ensures that the appropriate energy or momentum is deposited into the cells around the supernova, depending on whether the Sedov-Taylor phase of the blast wave is  resolved or not.  This prevents the supernova energy from being artificially radiated away, as would happen if solely thermal energy was injected (the so called over-cooling problem described in \citealt{Katzovercooling1992}). Both runs under study use cooling tables calculated by \cite{toocool4school}, down to $10^4$ K, and the \cite{Rosen1995} approximation for temperatures below this threshold.  A UV background is instantaneously turned on at $z=8.5$ to account for the re-ionisation of the Universe, while star formation is allowed to proceed when gas densities become greater than $4 \times 10^2 \rm{H.cm}^{-3}$ with an efficiency of 1\% per free-fall time, calibrated on observations by \cite{Kennicutt1998}.  A detailed description of the implementation of star formation used in this version of {\sc{ramses}}  may be found in \cite{Rasera2006} and \cite{Dubois2008}.  For the feedback run, a Chabrier initial mass function \cite{Chabrier2003} is adopted, with 31.7\% of the mass fraction of each star particle ending up as a single type II supernovae and releasing $10^{50} \, {\rm erg\, M}_{\odot }^{-1}$ of energy after a 10 Myr time delay and expelling heavy elements with a 5\% yield.

\begin{figure}
\begin{tikzpicture}
\begin{scope}[xshift=-1.5cm]
    \node[anchor=south west,inner sep=0] (image) at (0,0) {\includegraphics[width=0.9\columnwidth]{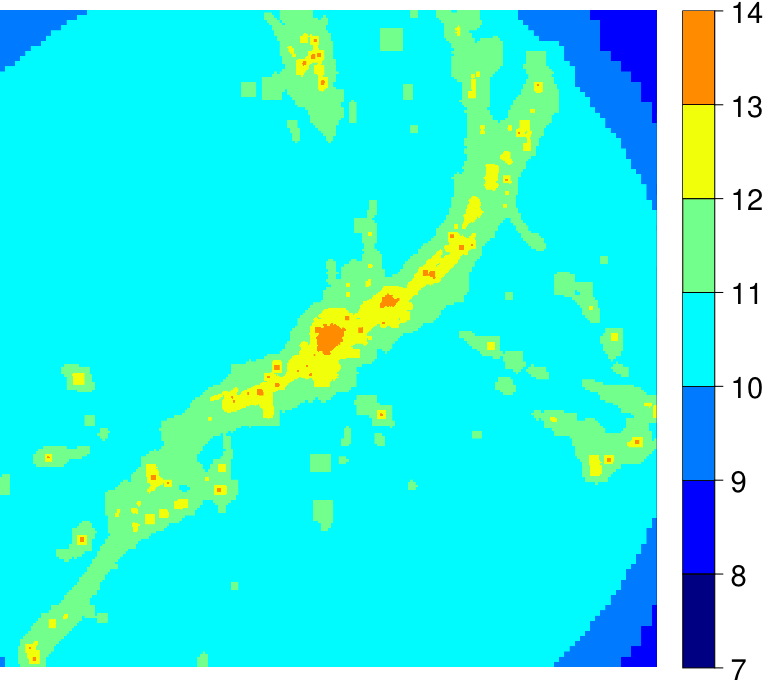}};
   
\end{scope}
\node [anchor=west] (note) at (6.05,3.5) {\Large Level};
\end{tikzpicture}
 \caption{Resolution map for a slice of thickness 300pc, across a (625kpc)$^2$ region of the computational domain at $z=4$, with each colour representing a different resolution level as indicated on the figure.
       At this redshift, the filament is uniformely sampled at 1.2 kpc resolution (AMR level 11: green) and partly at 0.61 kpc (AMR level 12: yellow) around the most massive halos embedded in it.  Even though the highest spatial resolution reached in the simulation is 10 pc, which corresponds to AMR level 20, levels above 13 are not shown as they are confined to the galaxies themselves and their immediate vicinity.}
    \label{fig:amr}
\end{figure}

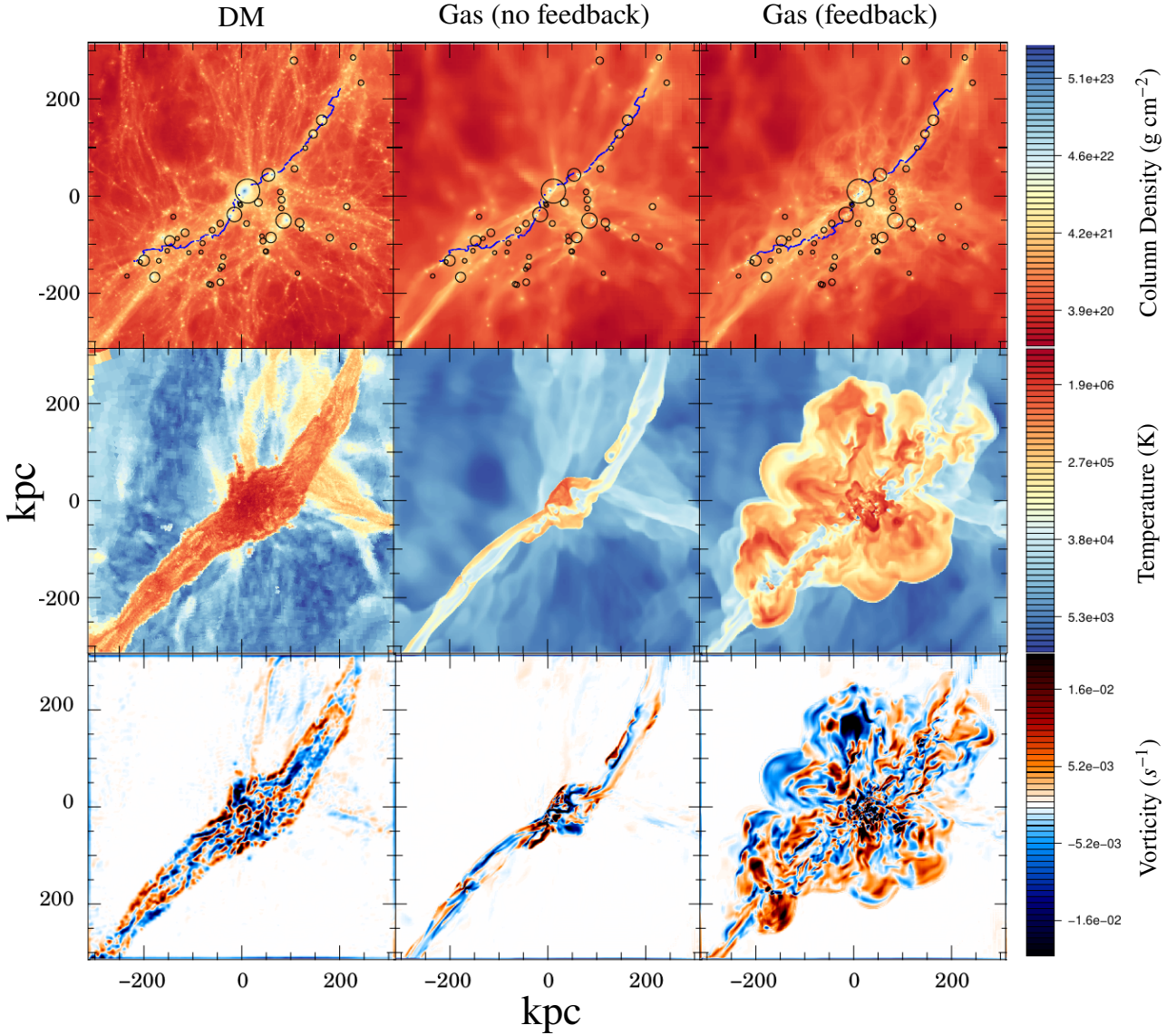
\begin{figure*}
\begin{tikzpicture}
\begin{scope}[xshift=-1.5cm]
    \node[anchor=south west,inner sep=0] (image) at (0,0) {\includegraphics[width=0.9\textwidth]{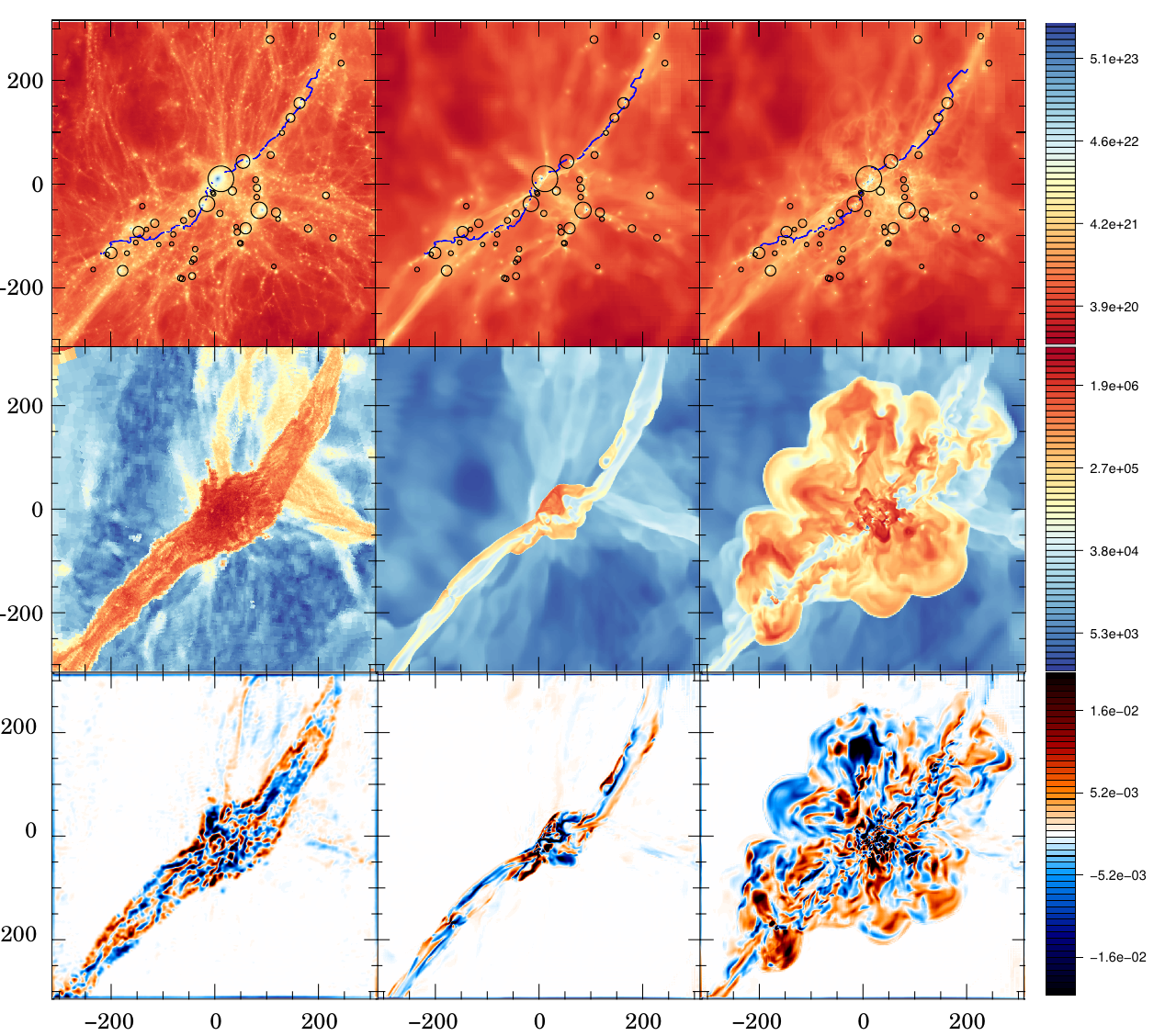}};
   
\end{scope}
\node [anchor=west,rotate=90] (note) at (-1.7,6.8) {\huge kpc};
\node [anchor=west] (note) at (5.4,-0.3) {\huge kpc};
\node [anchor=west] (note) at (1.,14.3) {\Large DM};
\node [anchor=west] (note) at (4.2,14.3) {\Large Gas (no feedback)};
\node [anchor=west] (note) at (8.9,14.3) {\Large Gas (feedback)};
\node [anchor=west,rotate=90] (note) at (14.6,9.8) {\large Column Density 
(g cm$^{-2}$)};
\node [anchor=west,rotate=90] (note) at (14.6,5.9) {\large Temperature 
(K)};
\node [anchor=west,rotate=90] (note) at (14.6,1.6) {\large Vorticity

($s^{-1}$)};

\end{tikzpicture}
     \caption{DM (left column) and gas (no-feedback run, middle column; feedback run, right column), with each row showing column density (\textit{top}), temperature (\textit{middle}) and vorticity (\textit{bottom}) in a slice 625 kpc across and 1 kpc thick at $z=4$. The main filament, as extracted from the DM density field, is overplotted ({\textit blue solid line}) on the column density maps. The virial radii of the 50 largest halos are marked as circles.  The differences between the feedback and no-feedback skeletons are caused by small differences in the noise level associated with DM  particles:  they yield slightly different paths which have a very similar length, so that either path can be chosen by the algorithm described in the text.  The colour bar for the density represents the gas.  To estimate it for the DM, one simply needs to divide the numbers shown by the universal baryon fraction. For the DM temperature, velocity dispersion is used as a proxy, with dark blue corresponding to regions of $\sim 0.02$ km s$^{-1}$ and deep red with $\sim 100$ km s$^{-1}$. In the vorticity panels, red represents matter swirling counter clockwise around the filament, and blue is for matter rotating in the opposite direction. 
     }
    \label{fig:merger}
\end{figure*}

\subsection{Filament Identification}
\label{sec:fil}
As we aim to measure the properties of the cosmic web filaments, both in the DM  and gas density fields, we now describe how we identify these structures in the simulations. 

\subsubsection{Method} The DM particle distribution is tessellated using the Delaunay Tessellation Field Estimator tool \citep{DTFE2007} and fed to the code {\sc DisPerSE} \citep{Sousbie11}.  {\sc DisPerSE} computes  stationary points (maxima, minima and saddle points) of the density field using the Hessian matrix and assigns to each pair of critical points (e.g. maxima-saddle) a persistence, namely a measure of how significant it is with respect to a Poisson distribution.  The persistence threshold is the single parameter that determines which features are considered as noise and which robustly pertain to the topology of the underlying density field. 
From this set of stationary points that characterize the topology of the field, {\sc DisPerSE} connects saddles to maxima following the direction of least gradient to create a network of filaments which will be referred to in this paper as the ``skeleton'', and we will call ``nodes" the maxima of the density field.

\subsubsection{Extraction of the skeleton from the simulations} 

For each simulation, filaments are extracted from the Delaunay tessellation reconstruction of the DM density field, setting a persistence threshold of 10$\sigma$. This persistence threshold is chosen such that the observed skeleton is in good visual agreement with the DM density field.  Our results are in fact insensitive to the exact value chosen for this threshold, as we are only studying the main filaments feeding the galaxy (see Section~\ref{ssec:CMF}). The skeleton is additionally processed with {\sc skelconv} (see {\sc DisPerSE} manual\footnote{\href{http://www2.iap.fr/users/sousbie/web/html/index4f3e.html?category/Manual}{http://www2.iap.fr/users/sousbie/web/html/index4f3e.html?category/Manual}})
using the {\sc breakdown} and {\sc smooth} functions.  {\sc breakdown} removes duplicate segments entering a node from two different starting points.  These segments can be so close as to be indistinguishable from one another and as such are removed to prevent their over-representation in the final skeleton. The skeleton is then smoothed by averaging over the positions of the 30 nearest 
neighbours of each segment.  This mitigates the effects of Poisson noise on the skeleton, ensuring that individual segments locally follow the global direction of the filament they belong to.\\
 In both the feedback and no-feedback runs, 1.22 physical kpc is the maximum spatial resolution reached in filaments, defined as the size of an individual cell on the highest AMR grid level that entirely maps the filament (see Fig.~\ref{fig:amr}). As is clear from Figure \ref{fig:amr}, higher refinement levels are triggered within filaments but their coverage is patchy, and mostly concentrated around halos/galaxies embedded within these elongated structures. As we argue in  our convergence analysis (Section~\ref{sec:Results}), we believe 1.22~kpc is enough to resolve the radial structure of filaments, at least those that connect to halos/galaxies with masses similar (or larger) to the one we study in this paper (roughly M$_\star$). 
We emphasize that this is a much higher resolution than that currently reached in large-scale cosmological hydrodynamics simulations, where $\simeq$ 1~kpc resolution is only attained within galaxies \citep[e.g.][]{Dubois2014, Vogelsberger2014,Nelson2018,DaveSimba2019,Schaye2015,Nelson2019TNG50}. The main drawback of our study is that such resolution is obtained at the cost of simulating a much smaller volume, and thus focuses on a single object.
Filament extraction is performed at the maximum level of resolution thus defined. However, as highlighted by \cite{Rosdahl2012} and in our convergence study (Section~\ref{sec:Results}), increasing the resolution does not seem to affect the filament properties much, and we thus expect that our results only weakly depend on resolution.  
\\
Finally, we note that {\sc DisPerSE} applied to the DM particle distribution, as is done in this paper, only allows the extraction of the filaments down to a scale comparable with the virial radius of DM halos.  Below this scale, DM filaments (at least in standard 3D space) are washed out by the virialisation process at the origin of halo formation and evolution. Therefore, we restrict our measurements of filament properties to filament segments located outside of the virial radii of embedded DM halos.

\begin{figure*}
\begin{tikzpicture}
\begin{scope}[xshift=-1.5cm]
    \node[anchor=south west,inner sep=0] (image) at (0,0) {\includegraphics[width=0.8\textwidth]{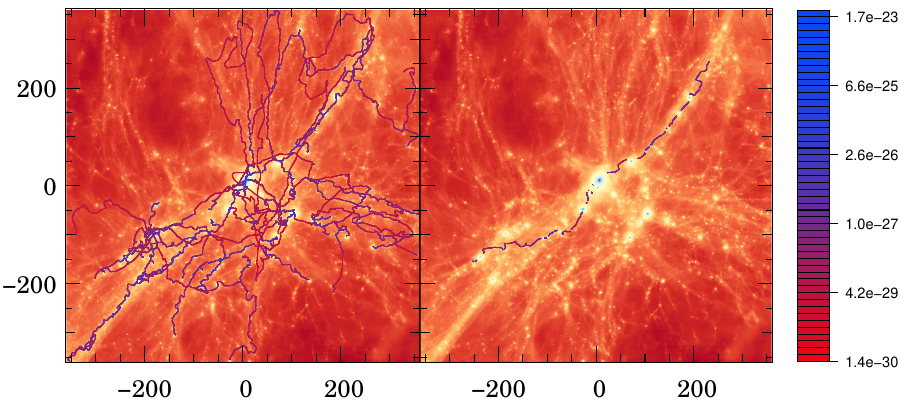}};
   
\end{scope}
\node [anchor=west,rotate=90] (note) at (-1.3,3.1) {\huge kpc};
\node [anchor=west] (note) at (4.6,-0.1) {\huge kpc};
\node [anchor=west,rotate=90] (note) at (13.,1.3) {\large Filament gas density (g/cm$^3$)};
\end{tikzpicture}
  \caption{The \textit{left} plot shows the raw skeleton extracted by {\sc DisPerSE}, which traces all the filaments of the DM density field, coloured according to the relative density (with low density in red and higher density in blue) . Using Dijkstra's algorithm we then obtain the skeleton on the right, where we have removed filament segments from regions with densities greater than 130 times the mean density, resulting in gaps around virialized halos and sub-halos (indicated by circles enclosing their virial radii on Fig.~\ref{fig:merger}).  In both panels, the skeleton is overplotted on a $z=4$ projection of the DM density field.}
    \label{fig:fils}
\end{figure*}
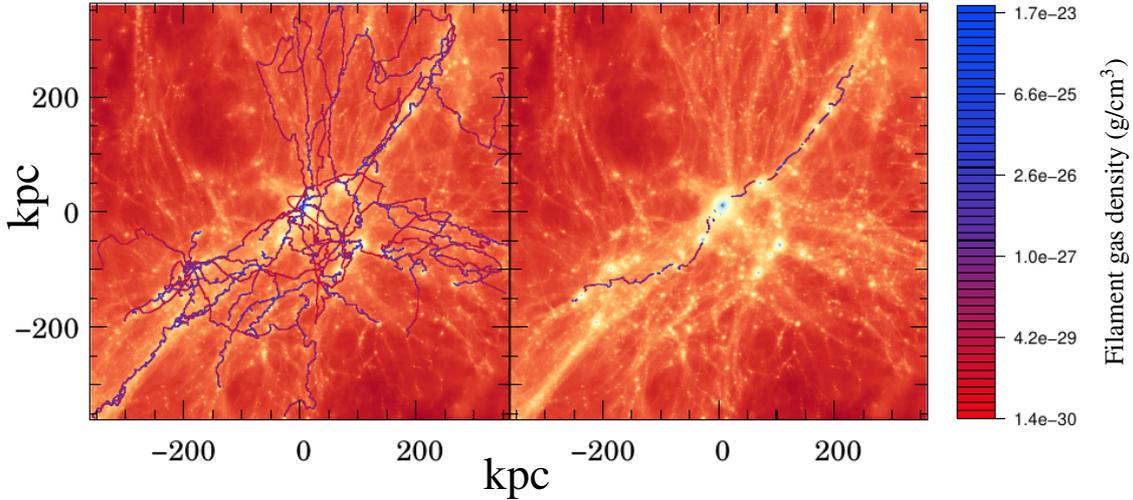

\subsubsection{Identifying the main filament} 
\label{ssec:CMF}
The gas and DM distributions differ significantly even for the no-feedback run (compare \textit{left} and \textit{middle} panels of Fig.~\ref{fig:merger}), with the gas density field presenting much fewer filamentary structures than the DM\footnote{This is a consequence of re-ionisation reheating the gas of the IGM and preventing accretion into the shallow potentials of DM filaments \citep{Katz2019}.}. In addition, even though dwarf galaxies residing within filaments are affected by feedback, the impact of this feedback on the growth of the central galaxy is minimal (it does not lead to the disruption of the main filament) as the majority of the gas feeding it at high redshift is accreted via  filaments, and not from mergers \citep{Danovich2012,Tillson15}. 
However, the gas density field in the run with feedback will be more perturbed due to 
interactions with galaxy winds and shocks (see \textit{middle} and \textit{right} panels of Fig.~\ref{fig:merger}), making the comparison between feedback and no-feedback runs  difficult. Furthermore, {\sc DisPerSE} is designed to work with particle data, as it allows in this case a meaningful definition of persistence (the very concept of which relies on quantifying the significance of a feature with respect to Poisson noise). 
For these reasons, and given that we are not interested in probing the existence of filaments within the virial radius of DM halos in this work, the DM density field seems more appropriate to carry out filament extraction.

We therefore elect to extract the skeleton from the DM density field, but trim it in order to keep only the main filament, along which most material flows onto the galaxy. For an M$_\star$ central galaxy, the main filament traced in the gas clearly coincides 
with its DM counterpart (see \textit{top} panels of Fig.~\ref{fig:merger}). As we are analyzing a filament connecting to a single object, we identify the approximate region where it begins and ends by eye, and select the highest density point in this region as its starting/end point\footnote{For larger volume cosmological simulations where an ensemble of filaments is available one can forgo the inspection by eye and simply use the closest pair of galaxies with similar masses which are linked by the skeleton as the starting and end points of a filament.}.  We then use Dijkstra's algorithm \citep{Dijkstra1959} to compute the shortest path (following the skeleton) between the start and end points. This works by assigning to each segment a distance from the start point, travelling along all the various possible paths of the skeleton.  Whenever a shorter path to a given segment, $s$, is found, then the selected path is updated up to $s$, and the distances to all segments connected to $s$ along this path which have a longer path length, are updated.  This process is iterated until the network is traversed, yielding the shortest path between the given start and end point.  The method is valid provided the main filament flows mostly straight onto the galaxy, which, in turn, holds until the filament gets close to the galaxy disk \citep{Powell11}.

\begin{figure*}
\begin{tikzpicture}
\begin{scope}[xshift=-1.5cm]
    \node[anchor=south west,inner sep=0] (image) at (0,0) {\includegraphics[width=0.9\textwidth]{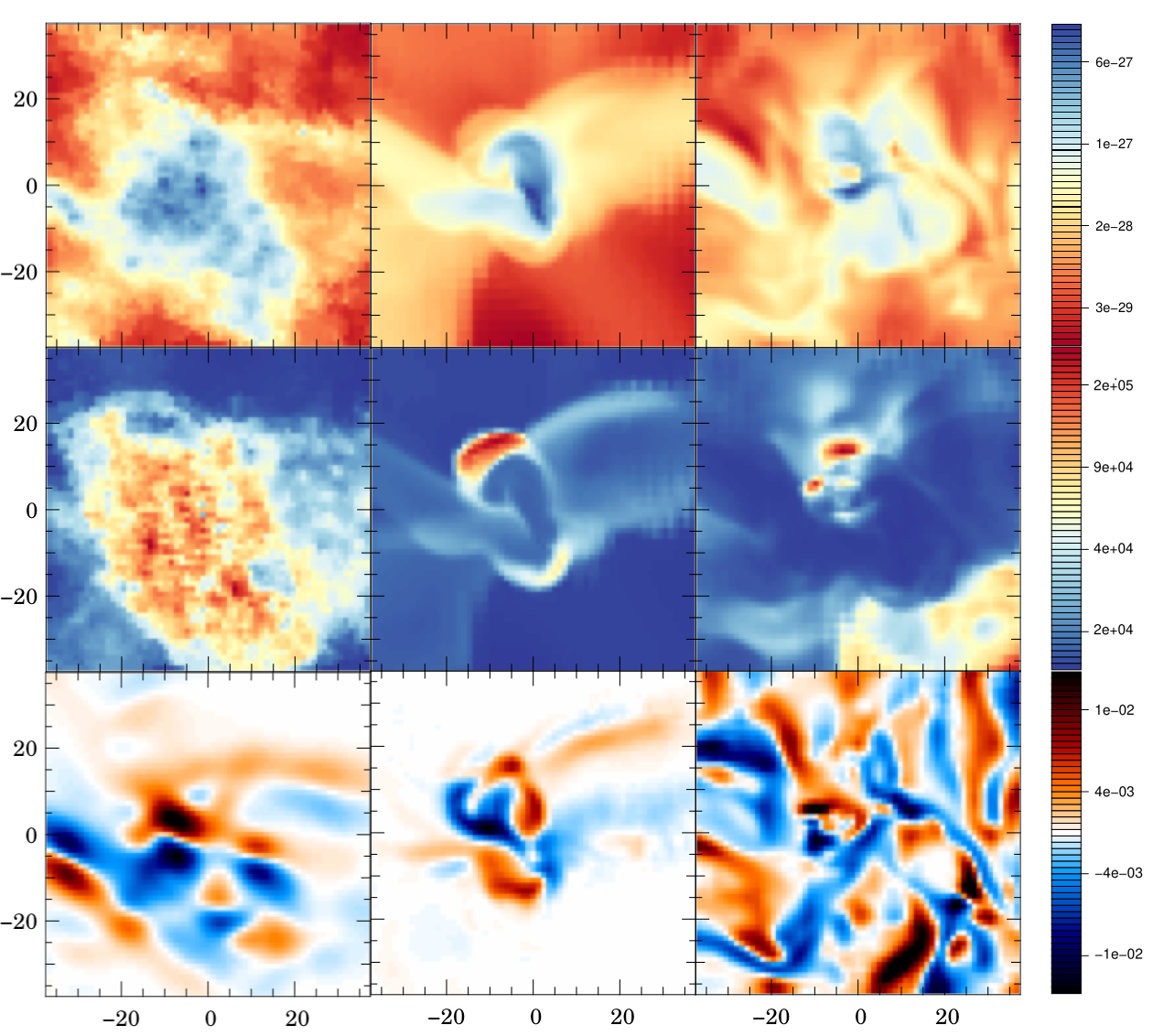}};
   
\end{scope}
\node [anchor=west,rotate=90] (note) at (-1.7,6.8) {\huge kpc};
\node [anchor=west] (note) at (5.4,-0.3) {\huge kpc};
\node [anchor=west] (note) at (1.,14.3) {\Large DM};
\node [anchor=west] (note) at (4.2,14.3) {\Large Gas (no feedback)};
\node [anchor=west] (note) at (9.,14.3) {\Large Gas (feedback)};
\node [anchor=west,rotate=90] (note) at (14.5,10.5) {\large Density 

(g cm$^{-3}$)};
\node [anchor=west,rotate=90] (note) at (14.5,5.9) {\large Temperature 

(K)};
\node [anchor=west,rotate=90] (note) at (14.5,1.6) {\large Vorticity 
(kms$^{-1}$ kpc$^{-1}$)};

\end{tikzpicture}
\caption{A typical filament cross-section, extracted 200~kpc away from the central galaxy in DM (left column) and gas (no-feedback run, middle column; feedback run, right column) at $z=4$. The thickness of the slice is of order 1 kpc. Note how the central filament (density peak in the 2D slice) is embedded in a weaker wall structure (which appears as a thick elongated tube encompassing the peak). From \textit{top} to \textit{bottom} row: density, temperature (or velocity dispersion for DM, running from 0 to 25 kms$^{-1}$, dark blue to red) and vorticity along the filament, with red representing matter rotating counter-clockwise and blue in the opposite direction.
    }
    \label{fig:planes}
\end{figure*}

In order to avoid the filament passing through halos, filament segments located in regions with densities higher than 130 times the mean density of the Universe were excluded\footnote{This value is lower than 200 times the critical density of the Universe which is commonly used in the literature to define virialised structures. This reflects the fact that the density of halos at the virial radius is lower than their average density by about a factor 3.}.  This density threshold is chosen empirically, but the resulting skeleton does not depend very sensitively on the chosen value provided this latter is on the order of 100 times the mean density of the Universe. The entire initial filamentary network and the resulting main filament extracted after post-processing are shown in Fig.~\ref{fig:fils}. Fig.~\ref{fig:merger} highlights that the skeletons extracted from the DM density fields of the feedback and no-feedback runs are slightly different. In this Figure, one can clearly see a pair of filaments on the left side of the central galaxy, which are in the final stages of merging.  As a result, our algorithm identifies two possible paths along which the main filament would have essentially the same length. Small changes in the noise level associated with the DM particles in the two different runs change the exact way that segments connect, resulting in the algorithm picking one of these paths in one run and the other path in the other run. Our results are, by and large, independent of such small randomly induced differences.

\subsection{Cross-section measurements} \label{ssec:Cross-Section}
\subsubsection{Calculating DM temperature and vorticity fields}

Due to the discrete Lagrangian nature of the numerical technique used to evolve the DM density field, a simple cloud-in-cell interpolation onto a reasonably sized regular grid generates a  non-smooth density field in poorly sampled, low density regions.  To get around this difficulty, a Delaunay tessellation \citep{SchappetVandeWeygaert2000} is computed from the DM density and velocity fields \citep[see e.g.][]{DTFE2007}, which ensures their spatial continuity.  The Delaunay grid is then projected onto a regular uniform grid, coinciding with AMR grid level 11, which corresponds to the maximum resolution mapping of the entire filament (cubic cells 1.22~kpc on a side, see Fig.~\ref{fig:amr}).  This uniform grid is used for measurement of all quantities in this paper unless otherwise stated.  The DM velocity dispersion field  -- used as a proxy for temperature -- is then obtained by computing the square of the difference between each particle velocity and the value of its nearest neighbour grid cell
and re-applying the Delaunay tessellation with this dispersion as the weight. Every time the Delaunay tessellation is projected onto the grid, we average all the tetrahedra (or volume fractions of) that co-exist in each grid cell.  The vorticity, on the other hand, is simply calculated by taking the curl of the velocity field on the uniform grid. As this latter is extremely noisy, a Gaussian smoothing is applied prior to computing vorticity, with a width of 2 cells.

\subsubsection{Cross-section extraction and radial profiles}
For each segment of the skeleton, a field (density, temperature or vorticity) is linearly interpolated in a plane, the thickness of which is equal to the skeleton segment length (typically 0.3~kpc, though this depends on the local density).  This plane is perpendicular to the segment and centred on it.  An example of individual cross-sections in the density, temperature and vorticity fields is displayed in Fig.~\ref{fig:planes}.  Note that the position of the DM or gas density peak does not necessarily lie exactly at the centre of the plane due to the smoothing of the skeleton.  Smoothing is required to ensure that individual segments point along the filament direction, and thus that the extracted planes are truly perpendicular to the filament.  The gas density maximum is not tied to the DM density maximum and thus is also unlikely to be at the centre of the plane. In order to correct for such small offsets which nevertheless do affect profile measurements, each plane is shifted using a method similar to the 'shrinking sphere' method outlined in \cite{Power2003}.  The centre of mass of a circle centred (with a radius greater than the truncation radius) on the initial guess from DisPerSE is calculated.  The circle is moved to the centre of mass before the procedure is repeated with a smaller circle.  This method is more robust to the presence of additional substructure within the filament, particularly as cells with $\rho>40\langle\rho\rangle$ have had their density reduced to $40\langle\rho\rangle$ for the calculation of the centre of mass.  This prevents halos existing within or near the filament from being chosen as the filament centre and distorting the filament profile.  DM and gas planes are therefore translated independently. This procedure allows us to align all segments when stacking cross-sections. 

Vorticity and temperature fields interpolated onto the plane perpendicular to the segments are then translated with the same shift as the density field.  
When looking in the plane perpendicular to them, filaments appear as strong peaks in the projected density field (see top row of Fig.~\ref{fig:planes}).  Alongside this, the major walls associated with these filaments is often visible extending out from the peaks, forming thick elongated structures which are not necessarily straight.  In the temperature field (middle row and middle column of Fig.~\ref{fig:planes}), strong radial shocks are observed around the filaments themselves, with weaker shocks also present at the wall
boundaries and where the walls intersect to form the filaments. In the vorticity field (bottom row, middle column of Fig.~\ref{fig:planes}) both filaments and walls are identified with the regions of highest vorticity amplitude. The DM filaments (left column of Figs.~\ref{fig:merger} and~\ref{fig:planes}) appear wider than their gaseous counterparts. Supernovae feedback (right column of Figs.~\ref{fig:merger} and~\ref{fig:planes}) renders filaments and walls imperceptible in the gas vorticity field (bottom right panels) although radial shocks are still present at the filament edges (middle right panel) and the gas density peak remains clearly visible (top right panel).

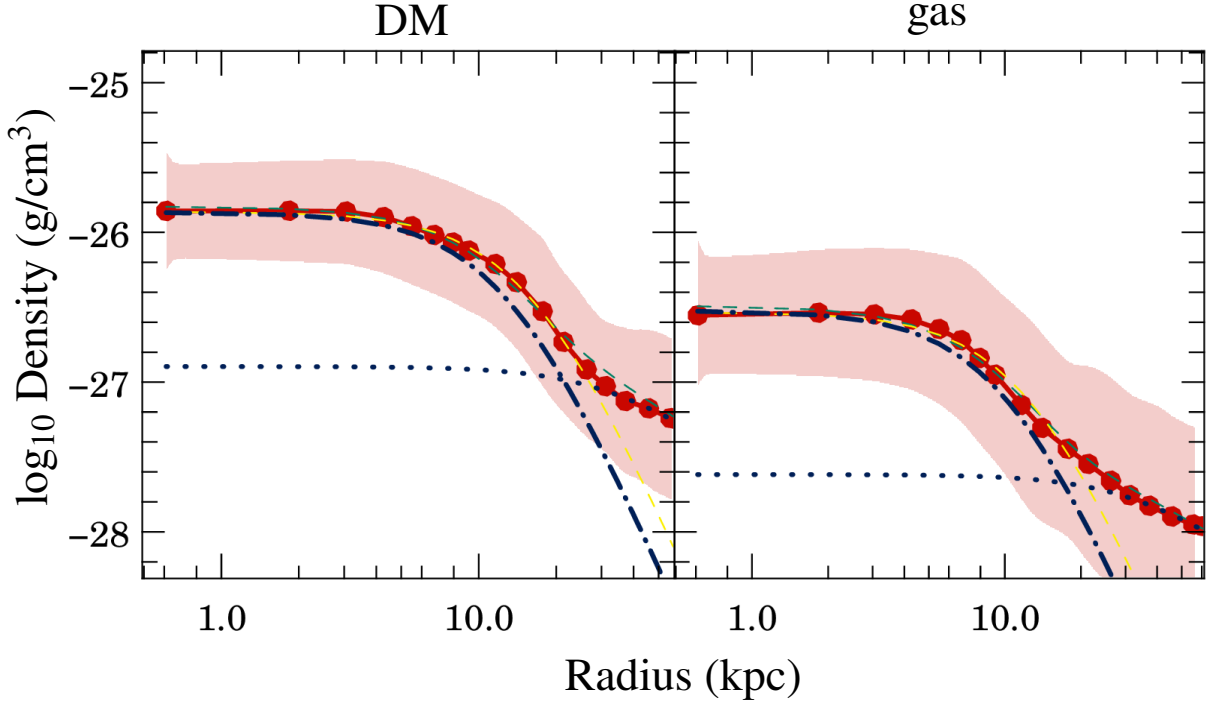
\begin{figure*}
\begin{tikzpicture}
\begin{scope}[xshift=-1.5cm]
    \node[anchor=south west,inner sep=0] (image) at (0,0) {\includegraphics[width=0.9\textwidth]{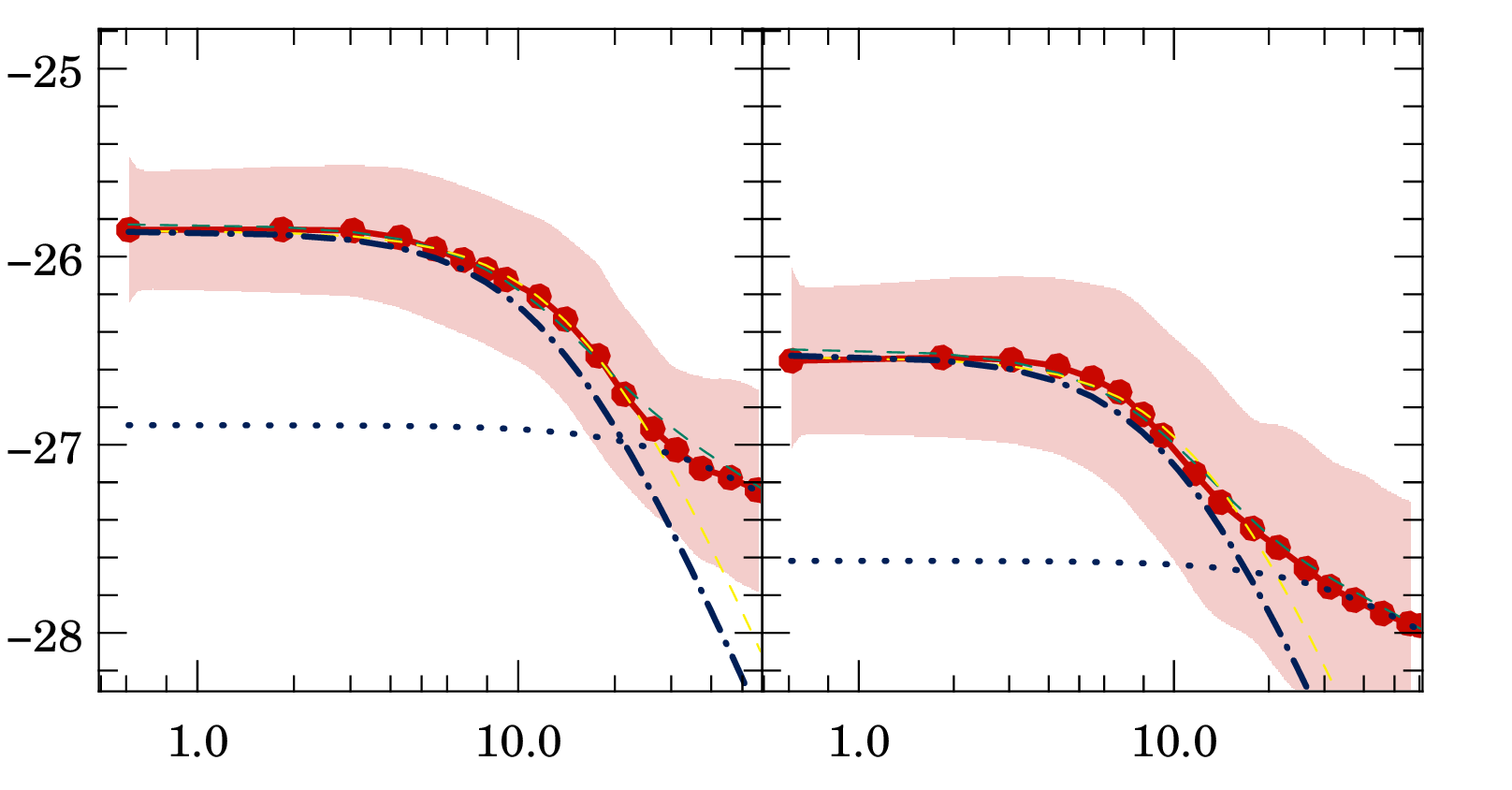}};
   
\end{scope}
\node [anchor=west,rotate=90] (note) at (-1.9,1.9) {\huge log$_{10}$ Density (g/cm$^3$)};
\node [anchor=west] (note) at (5.0,-0.3) {\huge Radius (kpc)};
\node [anchor=west] (note) at (2.5,8.45) {\huge DM};
\node [anchor=west] (note) at (9.5,8.45) {\huge gas};
\end{tikzpicture}
    \caption{Fits of the median DM (left panel) and gas (right panel) density profiles (red solid line and red disk symbols) at $z=4$, using a filament plus wall $\cal{M}_{\rm fil+wall}$ (green dashed line)
    or a filament only $\cal{M}_{\rm fil}$ (yellow dashed line) model.  The blue lines shows the $\cal{M}_{\rm fil+wall}$ decomposed in the filament (dot-dashed) and wall (dotted).  The shaded area  represents a 1 sigma deviation estimated by bootstrap re-sampling the values at each radial distance.
    }
   
    \label{fig:plummerfit}
\end{figure*}

 Radial profiles are measured from the 2D cross-sections by computing the azimuthal average in concentric shells centred on the highest density point.  
 When discussing the effects of resolution on the filament profile, we take the median value for the distribution of all filament segments at a given resolution and for each radius, as a single profile is required. However, for the rest of the measurements in this paper, we consider individual profiles fitted to each cross-section over the entire 
 radius range. 
 In Fig.~\ref{fig:plummerfit} the median profile obtained in that way \thesis{ is indicated by the filled red disk symbols joined by the red solid line, with the 1~$\sigma$ scatter around the median profile indicated by the shaded area.}  The advantage of this second method (fitting the whole profile) is that we can easily bin results according to other filament properties, such as distance to the central galaxy.  This should more accurately reflect the underlying filament property distribution. 

\subsubsection{The special case of vorticity cross-sections}
Vorticity being a vector, the structure of the vorticity field is far more complex than density or temperature (as illustrated by the bottom panels of Figs.~\ref{fig:merger} and ~\ref{fig:planes}) and, as a result, it not easy to stack individual vorticity profiles obtained for each skeleton segment. When stacking is required, we therefore use the modulus of the vorticity parallel to the direction of the filament and ignore azimuthal variations in vorticity.  The vorticity field in the direction of the filament is extracted in the same way as described in the previous section for the density and temperature.  As shown in the \textit{bottom} panels of Fig.~\ref{fig:planes}, the vorticity has a multipolar structure, with several rotating and counter-rotating vortices surrounding the filament. Outside the filament the amplitude of vorticity rapidly declines. Within filaments, the geometry of the vorticity field is mainly quadrupolar \citep[see][though we found that dipoles and higher order structures are not uncommon reflecting that the flow have shell-crossed several times]{Laigle15}.  This larger diversity in the structure of the vorticity field probably reflects the fact that the analysis in this paper looks at smaller scale vorticity than \cite{Laigle15}, and extends the measurement to gas. We recall that primordial vorticity is destroyed in an expanding Universe, and therefore voids are extremely vorticity poor. Vorticity can later be produced by shocks or shell-crossing (respectively for gas and DM), and as result is chiefly confined to walls, filaments and nodes \citep[see e.g.][]{Pichon1999}.  

\section{Results: measuring the filament profiles} 
\label{sec:Results}

In the following, we first derive analytically the radial profiles of filaments under the assumption that they are in hydrostatic equilibrium, and then compare them to the profiles directly measured in the simulation. 

\subsection{An analytic description of DM and gas filament profiles}

To obtain our analytic solution, we make the simple assumption that filaments may be modelled as infinite self-gravitating isothermal cylinders.  Fig.~\ref{fig:vel} presents the sound speed and velocity dispersion profiles in filaments.  We have been careful to subtract the bulk velocity of the material when extracting this data.  Within the filament, the sound speed and velocity dispersion are flat and dominate over the accretion velocity onto the filament, which suggests -- for the centre of the filament at least -- that the filament may indeed be treated as an isothermal cylinder in hydrostatic equilibrium\footnote{At least in the plane perpendicular to the filament, as we know that eventually,  
DM and gas flow along the filament into dark matter halos. In the steady state regime however, such a flow should not perturb the equilibrium.} , i.e.
\begin{equation}
\label{eq:hydrostat}
 \nabla \phi = - \frac{\nabla P}{\rho} \,,
\end{equation}
where $P=\mathcal{K}\rho$, and $\mathcal{K}={{\rm k}_{\rm B}T}/({\mu \, {\rm m}_{\rm p}})$, with $\rho$ the density, $T$ the temperature, ${\rm k}_{\rm B}$ the Boltzmann constant, ${\rm m}_{\rm p}$ the proton mass and $\mu$ the mean molecular weight of the gas.   \cite{Stodolkiewicz1963} solved this equation in the case of cylindrical symmetry \citep[see also][]{Ostriker1964}, and we will discuss the solution shortly. 
However, before we do, we briefly outline why it also applies to the collisionless DM fluid. Let us consider the time independent Jeans equations \citep{Jeans1915} for such a collisionless system: 
\begin{equation}
\label{eq:J2}
 \frac{\partial}{\partial x_i}(nv_i)=0\quad
 ;\quad
nv_i\frac{\partial v_j}{\partial x_i}=-n\frac{\partial \phi}{\partial x_j}-\frac{\partial (n\sigma_{ij}^2)}{\partial x_i}\,.
\end{equation}
where $v_i$ are the velocities, $\sigma_{ij}$ are velocity dispersions and $n$ is the DM number density. 
Under cylindrical symmetry, we may neglect all but the radial component of these equations. Further assuming steady state (i.e. that $v_r=0$, such that accretion onto the filament is negligible compared to internal pressure support) and that velocity dispersion is isotropic, the equations simplify to:
\begin{equation}
\label{eq:J3}
 \frac{d}{d r}(nv_r)=0\quad
 ;\quad
\frac{d \phi}{d r}=-\frac{1}{n}\frac{d (n\sigma^2)}{d r}\,.
\end{equation}
The second equation in (\ref{eq:J3}) is entirely analogous to equation (\ref{eq:hydrostat}), with $\mathcal{K}=\sigma^2$, though it is clear that accretion flow onto the filament cannot be ignored at large radii (see Fig.~\ref{fig:vel}).  The solution to both equations is therefore that given 
by \cite{Stodolkiewicz1963}:
\begin{equation}
\rho(r) = \frac{\rho_0}{\left(1+(r/r_0)^2\right)^2}\quad
{\rm with }\quad
r_0=\sqrt{\frac{2 \mathcal{K}}{\pi G \rho_0}}\,.
\label{eq:plummerprof}
\end{equation}
where $G$ is Newton's gravitational constant, $\rho_0$ is the central density and $r_0$ the core radius of the filament. Note however that it is the gravitational potential common to both components which should appear on the left hand side in both the second equation in (\ref{eq:J3}) and in eq.~(\ref{eq:hydrostat}), so that technically speaking only the DM is truly close to a self-gravitating isothermal cylinder, the gas being in hydrostatic equilibrium in the potential well of the DM filament.  

As a preliminary test of the model, we can use a typical DM filament central density of $\rho_0 \sim 1.7 \times 10^{-26} \mathrm{g \, cm}^{-3}$
, i.e. $\sim$100 times the mean density of the Universe (central filament density is subject to significant variations but this is typical of the median DM density, see left panel of Fig.~\ref{fig:plummerfit}) at $z=4$ for our choice of cosmological parameters, and a velocity dispersion $\sigma \sim 10\,$km/s (typical of the median DM velocity dispersion we measure, see Fig.~\ref{fig:vel}). Plugging these values in the second equation (\ref{eq:plummerprof}), we find a scale radius $r_0 \sim 8$kpc, which is broadly in agreement with the typical radius of the inner profile measured in the simulation, as shown in Fig.~\ref{fig:plummerfit} (left panel) and in table \ref{tbl:nofb}. As the width of the gas filament is set by the depth of the DM potential and the gas temperature, one needs to artificially use the DM central density for $\rho_0$ in equation (\ref{eq:plummerprof}) rather than that of the gas to obtain an estimate of $r_0$ for this latter (as shown on Fig~\ref{fig:plummerfit} the gas density is about a factor 5 lower than  that of the DM throughout the filament, in agreement with the universal value $\Omega_\mathrm{DM}/\Omega_B$). As the DM velocity dispersion for the particular filament we study is comparable to the gas sound speed (i.e. $\sim 10 \,\mathrm{km/s}$ or $10^4$K which corresponds the bottom temperature of the cooling curve for atomic hydrogen, see Fig.~\ref{fig:vel}) we expect a core radius for the gas similar to that of the DM, i.e. $r_0 \sim 8$kpc, and this indeed seems to be within a factor of 2 of the measured value (see right panel of Fig.~\ref{fig:plummerfit} and table \ref{tbl:nofb}).   

However, the isolated, infinite isothermal cylinder appears too highly idealised a model in at least one aspect, as can be seen by the failure of the yellow dashed curves (best fit obtained using the first equation~(\ref{eq:plummerprof})) to match the measured median profiles (solid red lines and red disk symbols) in Fig.~\ref{fig:plummerfit}.  In reality filaments are born from the intersection of walls (see Fig.~\ref{fig:planes} middle column panels), the presence of which will modify the filament profile, especially in the outer regions. Assuming that these walls may also be treated as hydrostatic atmospheres, but this time confined to a plane containing the filament, the equations governing their profiles are identical to eqs. (\ref{eq:hydrostat}) and (\ref{eq:J3}). These latter simply need to be solved in 1D instead of 2D, yielding in the direction $y$ perpendicular 
to the plane \citep[][]{Bookism1978}:
\begin{equation}
    \rho(y)=\rho_{y_0} \sech^2(y/h)\,,
        \label{eq:walldensprof}
\end{equation}
with the scale height $h = \sqrt{ \mathcal{K} / (2 \pi G \rho_{y_0})} $ taking a very similar functional form as $r_0$ in eq. (\ref{eq:plummerprof}), and $\rho_{y_0}$ standing for the density in the mid-plane ($y=0$) of the wall.  However, we need to integrate this wall profile over concentric cylindrical shells to evaluate how it modifies the filament profile. Unfortunately, this integral does not possess a simple analytic closed form, so we approximate the azimuthally averaged density of the wall by:
\begin{equation*}
    \rho(r)= \frac{\rho_{y_0}}{\alpha}  \frac{\tanh(\alpha r/h)}{r/h}\,,
\end{equation*}
with $\alpha=\pi/2$. Such an approximation captures the asymptotic behaviour of the correct solution for $r \gg h$, and is accurate to better than 14\% for all values of $r$.
As can be seen in Fig.~\ref{fig:plummerfit}, the inclusion of a wall modifies the shape of the outer filament. This might (at least partially) explain the discrepancy between filament density profiles previously reported in the literature, with power-law slopes ranging between -1 and -2 \citep[see e.g.][]{Colberg2005,Dolag2006,AragonCalvo2010}.  However we caution that these studies were performed on much larger scales and so may not be directly comparable to our work as they might potentially be affected by different biases.

One can easily show that in the case where the gas isothermal sound speed $c_s = \sqrt{{{\rm k}_{\rm B}T}/({\mu \, {\rm m}_{\rm p}})}$ equals the DM dispersion velocity $\sigma$, the density profiles of the gas and DM have the same exact shape, differing only by their normalisation, i.e. the value of the central density. In the more general case where these two velocities differ, one can write the gas density profile:
\begin{equation*}
     \rho_g(r)=\rho_{g_0}\left(1+ (r/r_0)^2\right)^{- 2 \sigma^2/c_s^2} \, ,
 \end{equation*}
so that if $c_s > \sigma$, it is shallower than that of the DM, and vice-versa. \cite{Katz2019} measured this effect by comparing two versions of the same cosmological simulation with and without reionisation.  They find that narrow streams are widened by the photo-heating of the gas, and that the gas counterparts of the lightest DM filaments can even be entirely erased. 

Note that this reasoning also applies to the isothermal gas density profile of a DM dominated isolated wall: when $c_s$ and $\sigma$ differ, it becomes  
\begin{equation*}
     \rho_g(y)=\rho_{gy_0}\sech^{ 2 \sigma^2/c_s^2} (y/h) \, ,
\label{eqn:wall}
 \end{equation*}

\subsection{Testing the simple model}
\label{sec:fit}
\begin{figure}
\begin{tikzpicture}
\begin{scope}[xshift=-1.5cm]
    \node[anchor=south west,inner sep=0] (image) at (0,0) {\includegraphics[width=0.9\columnwidth]{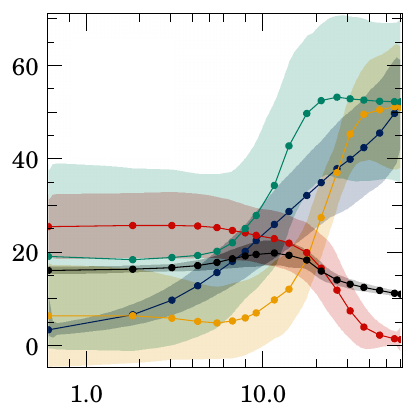}};
   
\end{scope}
\node [anchor=west,rotate=90] (note) at (-1.8,2.2) {\LARGE Velocity (kms$^{-1}$)};
\node [anchor=west] (note) at (1.4,-0.3) {\LARGE Radius (kpc)};
\node [anchor=west,text={rgb,255:red,1;green,1;blue,1}] (note) at (-0.1,6.7) {\Large c$_s$};
\node [anchor=west,text={rgb,255:red,0;green,130;blue,100}] (note) at (-0.1,6.4) {\Large v$_{r,g}$};
\node [anchor=west,text={rgb,255:red,202;green,10;blue,0}] (note) at (-0.1,6) {\Large $\sigma$};
\node [anchor=west,text={rgb,255:red,234;green,157;blue,0}] (note) at (-0.1,5.7) {\Large v$_{r,DM}$};
\node [anchor=west,text={rgb,255:red,0;green,30;blue,86}] (note) at (-0.1,5.4) {\Large v$_{c}$};

\end{tikzpicture}
  \caption{Median values for gas sound speed (black) from the no feedback run, DM velocity dispersion (red), gas (green) and DM (yellow) accretion velocity and circular velocity (blue) profiles at $z=4$, with shaded regions representing the 1 sigma scatter about the mode for each data point.  Note that in the inner filament region, one measures a near constant sound speed and velocity dispersion, which indicates the filament is, to a large extent, isothermal.  This breaks down at larger radii, due to both higher rates of radial inflow and falling sound speed and velocity dispersion.
    }
    \label{fig:vel}
\end{figure}
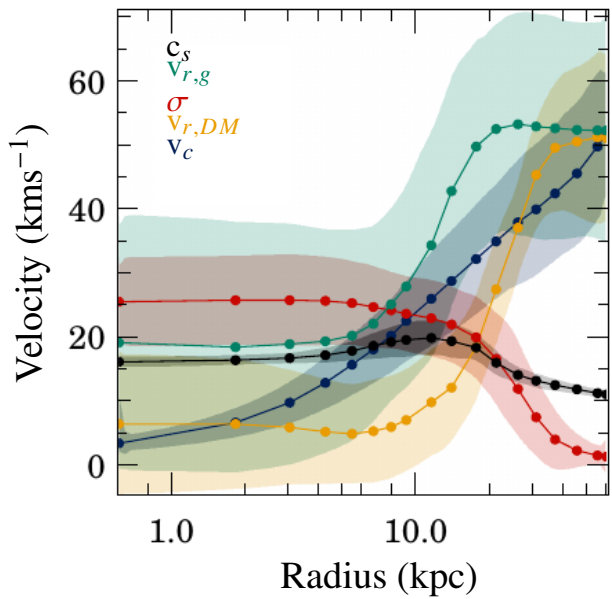

The first assumption we have made concerns the isothermality of the filament-wall system and that accretion onto the filament provides it with negligible support.  In Fig.~\ref{fig:vel} we can see that for the no feedback run both the gas sound speed (black solid line and solid disk symbols) and the DM velocity dispersion (red solid line and solid disk symbols) stay constant over most of the width of the filament, indicating that the isothermal approximation does indeed hold rather well.  In addition, the gas and DM accretion velocities are considerably lower than the sound speed and velocity dispersion respectively, and as such the dynamics of the system should be mainly driven by the pressure support. Moreover, the accretion shock itself is also non-adiabatic. Indeed, the upstream mach number (beyond 20 ~kpc) is $\mathcal{M}_u \approx 50/10=5$, thus the Rankine-Hugoniot jump conditions lead to a downstream Mach number $\mathcal{M}_d=0.47$, and it thus follows that the downstream sound speed (within 10 ~kpc of filament centre) should be $40 \,{\rm km\,s^{-1}}$, i.e. twice the value of that measured.  This indicates that the filament accretion shock is radiative rather than adiabatic.
Finally, we also plot on Fig.~\ref{fig:vel}, the circular velocity $V_c \equiv (GM/r)^{1/2}$ (blue solid line with solid disk symbols) measured for the filament, where $M$ is the mass enclosed by a cylindrical shell of radius $r$. We find that it is comparable to or lower than the sound speed/velocity dispersion, which further indicates that the filament is chiefly supported by pressure rather than by rotation, contrarily to what is argued in \citet{Mandelker2018}.

In Fig.~\ref{fig:plummerfit} we present 2 models, pure filament ($\cal{M}_{\rm fil}$), filament with wall ($\cal{M}_{\rm fil+wall}$). In practice, this means that along each individual skeleton segment, we fit the radial density using the formula:
\begin{equation}
     \rho(r)=\frac{\rho_0}{\left(1+ (r/r_0)^2\right)^2}+ \rho_1 \tanh(\alpha r/r_1) \frac{r_1}{r}\, ,
\label{eqn:fullfit}
 \end{equation}
where $r_1 = h$, and $\rho_1 = \rho_{y_0} \alpha^{-1}$. In principle, the values for $\sigma$ and $c_s$ could be different for the wall and embedded filament. However, for sake of simplicity and since we expect these two quantities to roughly behave in a similar manner, at least in the vicinity of the filament, we ignore the possible change in the ratio of $\sigma^2/c_s^2$ in our $\cal{M}_{\rm fil+wall}$ model (see Fig~\ref{fig:vel} for the validity of this assumption). The fit is performed using the Levenberg-Marquardt algorithm where the wall is first fit to the outer half of the profile with the filament contribution set to zero.  The wall parameters are then frozen in place while fitting the filament parameters.  In the case of a pure filament model (i.e. $\cal{M}_{\rm fil}$), the filament is fit to the entire profile, setting $\rho_1 = 0$.  The procedure was tested by applying it to a sample of artificial profiles, which typically returned radii within 1 cell of the input radius, but does break down when the filament is too wide (i.e. extends into the region where the wall is fitted).  While this is a suitable range for the purposes of this paper, filaments continue to grow as time progresses, and this method may become unsuitable at later times.  In Fig.~\ref{fig:plummerfit}, we show how each of the two models fares against the measured median density profile of the DM (left panel) or gas (right panel) at $z=4$.
Errors on the radius are estimated by considering the full distribution of density profiles measured from individual skeleton segments, and fitting this distribution with the best matched normal distribution to evaluate the value of the standard deviation.  For errors on the density, we use the best matched log-normal distribution instead, which is better suited to density distributions in filaments \citep[see e.g.][]{Cautun2014}.

Looking at Fig~\ref{fig:plummerfit}, it is not possible to distinguish the two models, $\cal{M}_{\rm fil+wall}$ (green curve) and $\cal{M}_{\rm fil}$ (yellow curve), in the inner region ($r \leq$20 kpc).  When the profiles are stacked as in this figure the fits work equally well with or without the wall. However the size of the core radius that these models return are very different when considering individual profiles: $r_0=19.34\pm7.15$~kpc for $\cal{M}_{\rm fil}$ compared to $r_0=8.39\pm3.82$~kpc for $\cal{M}_{\rm fil+wall}$ for the DM filament.  This factor of 2 discrepancy is also present for the gas filament: $r_0=8.99\pm1.86$~kpc for $\cal{M}_{\rm fil}$ compared to  $r_0=5.04\pm1.96$~kpc for $\cal{M}_{\rm fil+wall}$.  The core radii given by the $\cal{M}_{\rm fil}$ model can be rejected by simple visual inspection of Fig. \ref{fig:merger}: they are comparable to the outer edge radius of the filaments. 

Although we only show the $z=4$ median profiles, this behaviour of the two models holds for all redshift outputs examined in this work (see table~\ref{tbl:nofb} for a list). 

\begin{figure}
\begin{tikzpicture}
\begin{scope}[xshift=-1.5cm]
    \node[anchor=south west,inner sep=0] (image) at (0,0) {\includegraphics[width=0.9\columnwidth]{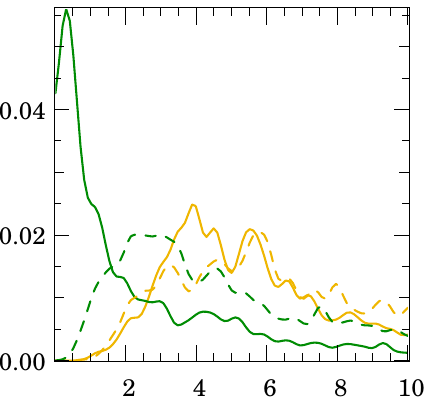}};   
\end{scope}
\node [anchor=west,rotate=90] (note) at (-1.8,3.5) {\LARGE p};
\node [anchor=west] (note) at (2.2,-0.3) {\LARGE $\chi^2_{\nu}$};
\end{tikzpicture}
 \caption{Normalised distributions of reduced $\chi_{\nu}^2$ obtained when fitting the filament only (yellow), or filament plus wall (green) models to filament density projections in individual slices perpendicular to each skeleton segment at $z=4$ (see text for detail). The gas filament is represented by the solid curves, whilst the DM counterpart is shown as dashed lines.}
    \label{fig:chi2s}
\end{figure}

We now discuss how each model fits individual cross-section density profiles (examples of these are shown as thin dashed curves on Fig.~\ref{fig:plummerfit}) rather than the median.
In this case, errors on the density are estimated by calculating the gradient of the density profile and multiplying it by the spatial resolution (size of cell). For the DM density, a Poisson noise contribution is also added in quadrature to the error estimate.  
We plot in Fig~\ref{fig:chi2s} the corresponding distributions of reduced $\chi_{\nu}^2$ which peaks at 3 for the DM density profile and 0.5 for the gas in the preferred model $\cal{M}_{\rm fil+wall}$ (dashed and solid green lines in Fig~\ref{fig:chi2s} respectively). For the $\cal{M}_{\rm fil}$ model these same distributions are much less strongly peaked around $\chi_\nu^2 = 6$ and $\chi_\nu^2 = 4$ for the DM and gas density profiles respectively (dashed and solid yellow curves).  Note that the measurement errors are relatively large, especially at the centre of the filament for the gas, and overall for the DM because of a significant Poisson noise contribution.  Though these values of $\chi^2$  suggest a fit to the simulation data which lies somewhat on the poor side, it is unclear that the validity of the model should be measured by $\chi^2$ statistics in the first place. Indeed individual profiles deviations from the model are very likely correlated with one another when substructures residing within the filament perturb its
density field. 

 For sake of completeness, let us mention that at $z=4$, the fits of the full set of skeleton segments using the $\cal{M}_{\rm fil+wall}$ model to the DM component for the no feedback run returns $r_0 = 8.39\pm3.82$~kpc as the mode and width of the fitted gaussian for the core radius of the filament (as previously mentioned), and $r_1=6.79\pm2.80$ kpc for the scale height of the wall.  Similarly we fit a log normal distribution to the DM central densities to obtain 
 $\log_{10}(\rho_0/{\rm gcm^{-3}})=-25.83\pm0.49$
 for the filament and
 $\log_{10}(\rho_1/{\rm gcm^{-3}})=-26.45\pm0.45$
 for the wall. As for the gas, we obtain $r_0 = 5.04\pm1.96$~kpc and $r_1=7.70\pm3.00$ kpc, with densities of $\log_{10}(\rho_0/{\rm gcm^{-3}})=-26.53\pm0.45$ and $\log_{10}(\rho_1/{\rm gcm^{-3}})=-27.12\pm0.37$.
A list of values for the filament radii and densities at other redshifts is provided in tables ~\ref{tbl:nofb} and ~\ref{tbl:nofbrho}.

As gas filament temperature remains around $10^4$~K at all times after re-ionization, their density profile flattens rapidly as the mass of their DM counterpart decreases and the sound speed approaches the critical value of $c_s=\sqrt{2}\sigma$. This means that low mass filaments will only exist in the DM component (compare the top left and middle panels in Fig~\ref{fig:merger}), as a $10^4$ K gas has too much pressure to be trapped in the DM potential well in that case, and thus, talking about a gas $r_0$ becomes quite meaningless. On the other side of the mass range, we expect more massive filaments, where DM has a larger velocity dispersion, to have better defined cores in the gas than DM, as this former should still radiatively cool down to $\sim 10^4$~K and thus have a much steeper density profile than its DM counterpart. As a result of this cooling, it is possible that the central gas density of massive filaments will become comparable to that of the DM, in which case our assumption that the DM sets the gravitational potential would cease to be valid and the core radii of the two components might then differ substantially. However, for the filament system considered in this paper, the approximation of similar DM and gas density profiles seems to hold quite well (see Fig~\ref{fig:plummerfit}).

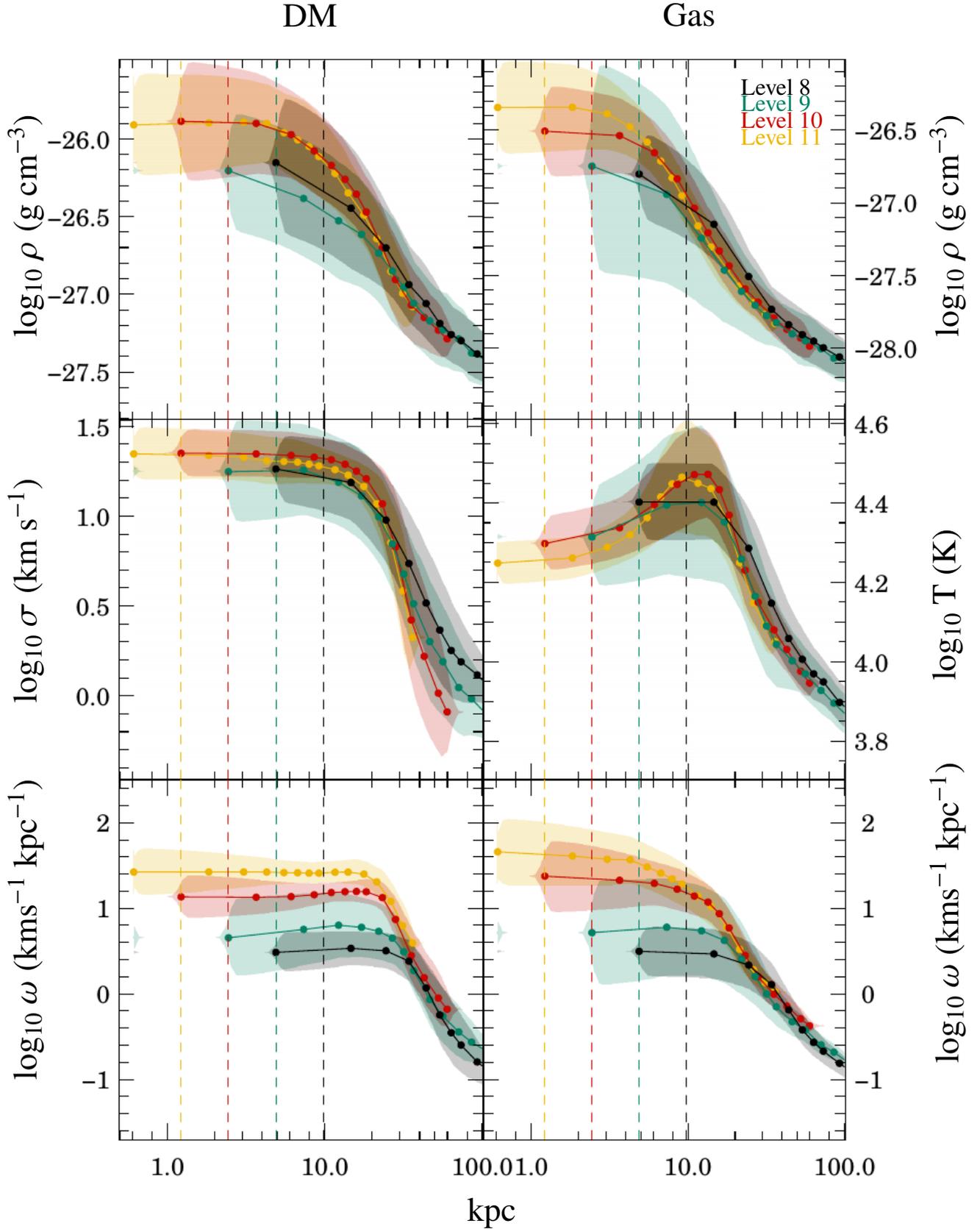
\begin{figure*}
\begin{tikzpicture}
\begin{scope}[xshift=-1.5cm]
    \node[anchor=south west,inner sep=0] (image) at (0,0) {\includegraphics[width=0.9\textwidth]{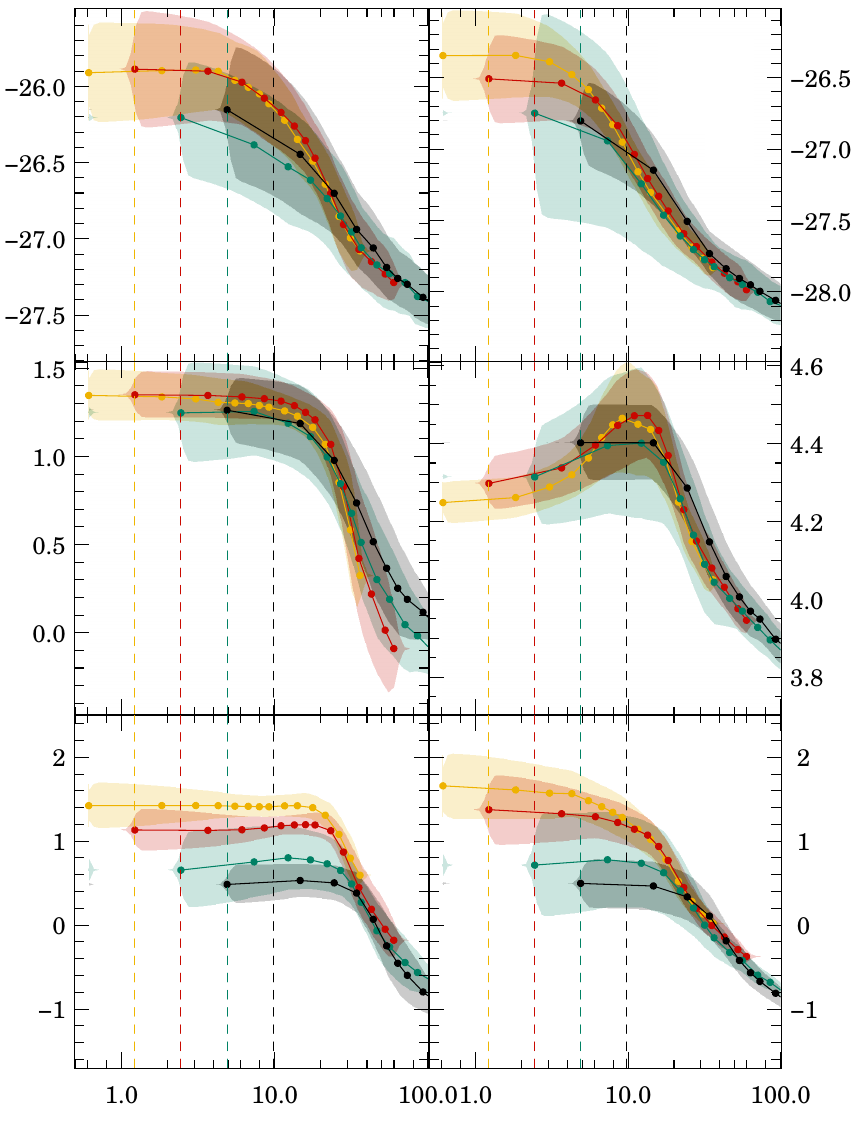}};
   
\end{scope}
\node [anchor=west,rotate=90] (note) at (-1.8,15.4) {\huge log$_{10}\:\rho$ (g cm$^{-3}$)};
\node [anchor=west,rotate=90] (note) at (-1.8,8.8) {\huge log$_{10}\:\sigma$ (km s$^{-1}$)};
\node [anchor=west,rotate=90] (note) at (-1.8,2) {\huge log$_{10}\:\omega$ (kms$^{-1}$ kpc$^{-1}$)};
\node [anchor=west] (note) at (6,-0.3) {\huge kpc};
\node [anchor=west,rotate=90] (note) at (14.7,15.4) {\huge log$_{10}\:\rho$ (g cm$^{-3}$)};
\node [anchor=west,rotate=90] (note) at (14.7,9.2) {\huge log$_{10}\:$T (K)};
\node [anchor=west,rotate=90] (note) at (14.7,2.3) {\huge log$_{10}\:\omega$ (kms$^{-1}$ kpc$^{-1}$)};

\node [anchor=west] (note) at (2.7,21.3) {\huge DM};
\node [anchor=west] (note) at (9.5,21.3) {\huge Gas};

\node [anchor=west,text={rgb,255:red,1;green,1;blue,1}] (note) at (10.9,20) {\Large Level 8};
\node [anchor=west,text={rgb,255:red,0;green,130;blue,100}] (note) at (10.9,19.7) {\Large Level 9};
\node [anchor=west,text={rgb,255:red,202;green,10;blue,0}] (note) at (10.9,19.4) {\Large Level 10};
\node [anchor=west,text={rgb,255:red,239;green,182;blue,0}] (note) at (10.9,19.1) {\Large Level 11};

\end{tikzpicture}
  \caption{The median radial profiles of the filaments in DM (left column) and gas in the no-feedback run (right column), for density (top row), temperature (middle row) and vorticity  (bottom row) at $z=4$. Displayed profiles (from black to yellow) represent data extracted at different spatial resolutions of the AMR simulation grid (vertical dashed lines or levels 8 to 11 respectively, see text for detail). Error bars are generated by bootstrapping the distribution of individual filaments profiles, and taking the root mean square at each radius.}
    \label{fig:profiles}
\end{figure*}

In light of the previous discussion, we interpret the difference between the measured and predicted median values of $r_0$ as a departure from the isothermal/hydrostatic approximations for the filament (see Fig.~\ref{fig:profiles}, middle panels), 
rather than to asymmetry or a systematic variation of core size as a function of distance to the galaxy (see section~\ref{fig:ppp} for more detail concerning this latter variation). For the DM, the filament median velocity dispersion varies by 10\% within $\sim 2-3$ core radii. For the gas, where a shock is clearly visible around 15~kpc away from the centre of the filament (middle right panel of Fig.~\ref{fig:profiles}), the temperature varies by less than 60\% between the centre of the filament and the maximum of the shock.  

These discrepancies notwithstanding, it is striking how filaments in our cosmological simulations resemble those obtained in a much more idealised set-up with similar resolution presented in \cite{Klar2012}. More specifically, even though these authors ignore the DM component as well as the fragmentation and mergers of filaments, they find that their gaseous linear structures are in radial hydrostatic equilibrium and exhibit an isothermal core several kpc wide with central densities and temperatures remarkably similar to those we measure in a more realistic context. They also identify an outer shocked region with similar properties as ours, but with a gravitational focusing which reduces $r_0$ and increases $\rho_0$ as the filament approaches the DM halo to which it is connected. As previously mentioned, we will come back to this latter point in section~\ref{sec:temp} of our results devoted to the temporal and spatial evolution of filament properties, but already note that such a focusing effect is not as pronounced in our simulations.

\subsection{Vorticity, temperature and the radial extent of filaments}

Having extracted the main filament from the simulation and measured the characteristic radius of its core through the use of a simplified model of hydrostatic equilibrium for its density profile, we now turn to the question of determining its outer size, or \textit{truncation} radius, as the analytic profile cannot extend to infinity in the radial direction.

Beyond a certain radius it is no longer true that sound and dispersion velocities dominate over the accretion velocity, as can be seen in Fig. \ref{fig:vel}. A failure of the hydrostatic model will thus occur, leading to a potential definition of the truncation radius, which also coincides with the position of the accretion shock onto the filament for the gas. We have opted not to use the peak temperature position as a definition of the truncation radius, as individual skeleton segment profiles, both in temperature and vorticity are often asymmetric and/or distorted by their environment, and may contain multiple peaks when averaged over concentric radial shells as a result (see Fig~\ref{fig:planes} for an example). Moreover, 
such a definition would not apply to DM velocity dispersion profiles. We have therefore chosen to use a universal method for all physical quantities and types of filament (gas or DM), which also has the benefit of providing internal consistency between measurements. We 
thus define the truncation radius as the point where the steepest descent in the temperature/vorticity/velocity dispersion profile is attained.  \corrections{In the DM, this is analogous to the splashback radius for halos as defined in \cite{Diemer2017}, as vorticity and velocity dispersion is only generated in the DM where shell crossing has occurred}  At $z=4$ this yields a truncation radius of $18.6\pm4.0$ kpc for the gas temperature profile, while the gas vorticity profile gives $21.7\pm6.6$ kpc. The DM filament at the same redshift has a measured truncation radius of $28.6\pm6.5$ kpc when using the velocity dispersion profile, and $25.9\pm4.5$ kpc if we consider its vorticity profile (see Table~\ref{tbl:nofb}).  It is interesting to note that the accretion shock of the gas filament seems positioned well within the DM filament (roughly at half the DM truncation radius).
In order to check the robustness of these measurements vis-\`a-vis resolution the data was extracted at four different spatial resolutions. Note that this is \textit{not} a study where we change the resolution and re-run the simulation, but simply a post-processing of the \textit{same} simulation at different resolutions, so we expect to achieve better agreement than if we had done a proper resolution study. As resolution increases progressively from 10 kpc to 1.22 kpc (level 8 to 11), the profiles are seen to converge across every panel of Fig~\ref{fig:profiles}. Note that for comparison, our \textit{lowest} level of resolution, i.e. 10 kpc roughly corresponds to the \textit{highest} level of resolution available to capture filaments in current cosmological simulations with volumes on the order of 100 Mpc on the side, like Mare Nostrum \citep{Ocvirk2008}, Horizon-AGN \citep{Dubois2014}, MassiveBlackII \citep{Khandai2015}, Eagle \citep{Schaye2015}, IllustrisTNG \citep{Nelson2018} or SIMBA \citep{DaveSimba2019}.

Looking first at the density profiles both of the DM and gas filaments (top panels of Fig.~\ref{fig:profiles}), one can see that in going from the highest resolution level to the lowest one, the central density (inside the core) is underestimated by about an order of magnitude, and one becomes unable to measure the core radius of the profile with a reasonable accuracy.
On the other hand, the DM velocity dispersion profiles (middle left panel of Fig~\ref{fig:profiles}) seem to converge faster than the density ones, with the lower resolution estimates compatible with the higher resolution ones at all radii. 
This seemingly rapid convergence is induced by the shape of the isothermal profiles which are, by definition, flat, especially in the case of the DM. For the gas (middle right panel of Fig~\ref{fig:profiles}), the temperature does not show as marked a convergence as the DM velocity dispersion because of the presence of the accretion shock: the low resolution data (black curve), which barely resolves the truncation radius of the filament underestimates the shock temperature and overestimates the core temperature by a similar amount.     
Having said that, the  shock position is fairly robust to resolution changes despite being radially asymmetric, which leads to its `smearing'. A minimum resolution of 2.4 kpc is required to correctly capture both the temperature of the accretion shock and that of the gas filament core.

\subsection{Evolution of filament profiles over cosmic time and distance from central halo}
\label{sec:temp}

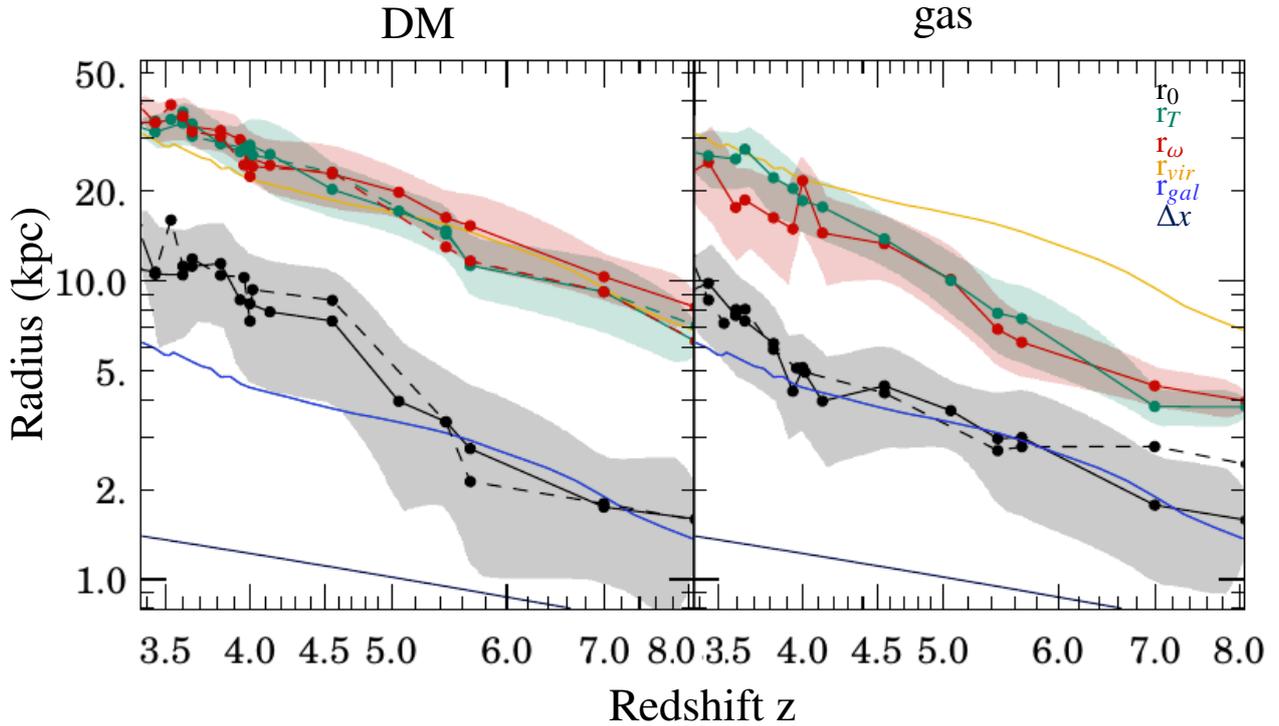
\begin{figure*}
\begin{tikzpicture}
\begin{scope}[xshift=-1.5cm]
    \node[anchor=south west,inner sep=0] (image) at (0,0) {\includegraphics[width=0.9\textwidth]{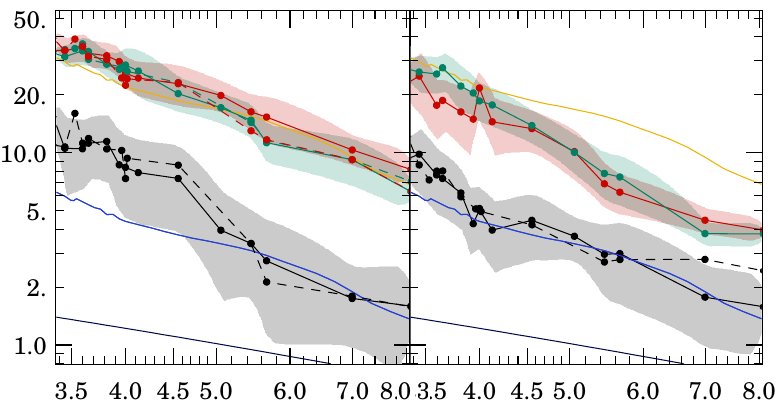}};
   
\end{scope}
\node [anchor=west,rotate=90] (note) at (-1.8,3) {\huge Radius (kpc)};
\node [anchor=west] (note) at (5.7,-0.4) {\huge Redshift z};
\node [anchor=west] (note) at (2.7,8.65) {\huge DM};
\node [anchor=west] (note) at (9.7,8.65) {\huge gas};
\node [anchor=west,text={rgb,255:red,1;green,1;blue,1}] (note) at (12.9,7.7) {\Large r$_0$};
\node [anchor=west,text={rgb,255:red,0;green,130;blue,100}] (note) at (12.9,7.4) {\Large r$_{T}$};
\node [anchor=west,text={rgb,255:red,202;green,10;blue,0}] (note) at (12.9,7) {\Large r$_{\omega}$};
\node [anchor=west,text={rgb,255:red,234;green,157;blue,0}] (note) at (12.9,6.7) {\Large r$_{vir}$};
\node [anchor=west,text={rgb,255:red,0;green,30;blue,86}] (note) at (12.9,6.1) {\Large $\Delta x$};
\node [anchor=west,text={rgb,255:red,50;green,50;blue,255}] (note) at (12.9,6.4) {\Large r$_{gal}$};

\end{tikzpicture}
       \caption{
    Evolution of the core (black curves) and truncation radii (red and green curves for estimates based on the vorticity and temperature respectively) of the filaments with redshift. The left panel represents the DM filament, and its gas counterpart is on the right.  Dashed lines represent the feedback run. The virial radius is shown in orange and the approximate extent of the central galaxy (20\% of virial radius) in yellow. Finally, the spatial resolution of the simulation in the filament is indicated by the solid blue line at the bottom of each panel.
    }
    \label{fig:radius}
\end{figure*}

Having focused, so far, the discussion of the filament profile at $z=4$, we now address the issue of its temporal evolution. Since  $r_0= \sqrt{{2 \mathcal{K}}/{\pi G \rho_0}}$, we naively expect that $r_0  \propto  (1+z)^{-3/2} $, provided the filament central density scales with that of the background Universe --- which we measure to be the case (see Table~\ref{tbl:nofbrho}) --- and its central temperature/velocity dispersion remains roughly constant with redshift. Conversely, we can deduce the scaling of filament temperature/velocity dispersion with redshift by measuring the departure of $r_0$ from this specific power law scaling. In our simulation, we find that for the gaseous filament, the central radius grows as $r_0 \propto (1+z)^{\rono\pm\ronoerr}$, which means that the sound speed should scale like $c_s \propto (1+z)^{-1.22\pm0.12}$
whereas we measure $c_s \propto (1+z)^{-1.41\pm0.28}$, i.e. an evolution quite consistent with the naive expectation. 

For the DM filament counterpart, the growth of $r_0$ is faster, with a measurement of $ r_0 \propto (1+z)^{\ronoDM\pm\ronoDMerr}$ (see Fig~\ref{fig:radius}), a faster rate than the approximate size of the central galaxy ($r_{\rm gal} = 0.2 r_{\rm vir}$, blue solid line on the Figure). This implies that $\sigma \propto (1+z)^{-1.68\pm0.15}$ as redshift decreases, whereas we measure in the simulation that $\sigma$ scales as $(1+z)^{-1.46\pm0.39}$.  The evolution of both gas and DM filament core radii are therefore consistent with the naive expectation at a $\sim 1$ $\sigma$ confidence level.  \thesis{Strictly speaking, any evolution of the central sound speed and velocity dispersion is in contradiction with the underlying assumption of isothermality used to derive the filament profiles, as this latter requires {\it no} change in either quantity with redshift.  However, as the evolution is slow compared to the sound crossing time of the central region, an instantaneous isothermal profile fits the data fairly well.}

The explanation for the somewhat faster growth of the core radius of the filaments than the radius of the central halo to which it is connected is that the 'old' core material is preferentially drained by halos residing within the filament, while a 'new' core forms out of more freshly accreted matter onto the filament \cite[see e.g.][]{Pichon11}. As a result, the filament core radius is more sensitive to the recent accretion history onto the filament than the halo.  Such a behaviour is reminiscent, at least qualitatively, to that of the Navarro--Frenk--White density profile scale radius, $r_s$,  found by e.g. \citet{Munoz2011} whose time evolution also differs significantly from that of the virial radius of the DM halo (except in that case it is the opposite: $r_s$, which is less sensitive to the halo recent accretion history, starts decreasing with redshift earlier than $r_{\rm vir}$, see their Figure 5). As the gas can be considered, to first order, in hydrostatic equilibrium in the DM filament potential well, we expect the evolution of its core radius to be somewhat influenced by that of the DM, i.e. that its growth also be sped up. We intend to explore this effect in more detail and with a larger sample of filaments  to better assess the universality of this behaviour.

 As for the truncation radius for the gas/DM filaments, determined from either the vorticity or the temperature/velocity dispersion, it represents the locus where fresh material is accreting, and as such is the rough equivalent of the halo virial radius.  Fig.~\ref{fig:radius} shows the evolution of this radius as a function of redshift, along with the size of the main halo embedded in the filament ($r_{\rm vir}$, orange solid line).  
 For the DM filament, the truncation radius evolves as  $r_\omega  \propto  (1+z)^{\rwnoDM\pm\rwnoDMerr}$ or $r_T  \propto  (1+z)^{\rTnoDM\pm\rTnoDMerr}$, depending on whether ones uses vorticity or velocity dispersion to define it. This is a growth rate very similar to that of the halo size $r_{\rm vir} \propto (1+z)^{-2.11\pm0.02}$ in this range of redshifts.  However, the gas truncation radius, either derived from the vorticity or temperature of the gas filament which scale as $r_\omega  \propto  (1+z)^{\rwno\pm\rwnoerr}$ and 
 $r_T  \propto  (1+z)^{\rTno\pm\rTnoerr}$ respectively, grows significantly faster than its DM counterpart. This is reminiscent of the stability driven argument for the propagation of a radiative shock within DM halos advanced by \citet{Birnboim2003}, but this time applied to the filament: as time progresses and density drops the shock is able to propagate outwards and ends up filling the entire DM filament volume. Practically, 
 this means that even though the gas filament starts off being smaller than the central halo embedded within it (see Fig.~\ref{fig:radius}) at high redshift, the truncation radius rapidly catches up with the virial radius. In our specific case, they are essentially the same size by $z=3.5$.

\begin{figure}
\begin{tikzpicture}
\begin{scope}[xshift=-1.5cm]
    \node[anchor=south west,inner sep=0] (image) at (0,0) {\includegraphics[width=0.9\columnwidth]{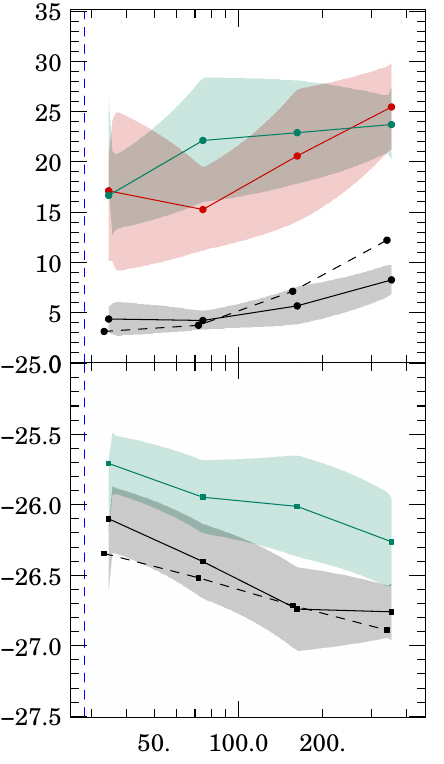}};
   
\end{scope}
\node [anchor=west,rotate=90] (note) at (-2,8.9) {\Large Radius (kpc)};
\node [anchor=west,rotate=90] (note) at (-2,2.3) {\Large log$_{10}\:\rho$ (g/cm$^3$)};
\node [anchor=west] (note) at (-0.4,-0.4) {\Large Distance to central galaxy (kpc)};
\
\node [anchor=west,text={rgb,255:red,1;green,1;blue,1}] (note) at (0.1,12.7) {\Large r$_0$};
\node [anchor=west,text={rgb,255:red,0;green,130;blue,100}] (note) at (0.1,12.4) {\Large r$_{T}$};
\node [anchor=west,text={rgb,255:red,202;green,10;blue,0}] (note) at (0.1,12.1) {\Large r$_{\omega}$};
\node [anchor=west,text={rgb,255:red,0;green,30;blue,86}] (note) at (0.1,11.8) {\Large r$_{vir}$};
\node [anchor=west,text={rgb,255:red,1;green,1;blue,1}] (note) at (0.1,6.3) {\Large $\rho_g$};
\node [anchor=west,text={rgb,255:red,0;green,130;blue,100}] (note) at (0.1,6) {\Large $\rho_{DM}$};

\end{tikzpicture}
  \caption{\textit{Top:} Gas filament radii as a function of distance to the galaxy, in the no-feedback  (\textit{solid} line) and feedback runs (\textit{dashed} line) at $z=4$. The core radius is in black, whereas the truncation radius estimated from the temperature and the vorticity profiles are in green and red respectively. Note that the truncation radius cannot be determined from either the vorticity or temperature when feedback is included. 
   \textit{Bottom:} Central density of the filament as a function of distance to the galaxy.  Gas is black, DM is green, with solid and dashed lines representing no feedback and feedback runs respectively. The vertical blue dashed line indicates the virial radius of the halo.}
    \label{fig:ppp}   
\end{figure}

We now go back to $z=4$ to explore the effect of distance to the galaxy on the width of the filament.  As can be seen in Fig. \ref{fig:ppp} (top panel), both the core and  truncation radii of the gas filament decrease by less than a factor 2 as a function of the distance to the main galaxy embedded within it. This decrease is progressive, from a maximum radius at $~ 300$ kpc away, which corresponds to the distance of either ends of the filament (see Fig. \ref{fig:zoom}), to the virial radius of the central galaxy. We caution the reader that this is somewhat different to the reported behaviour of the filament once it enters the virial radius of the embedded DM halo \citep[e.g.][]{Danovich2012}. Indeed, within the virial radius, one expects the gas filament to undergo more important gravitational focusing \cite[][]{Klar2012}. The reason why this does not happen as strongly in our case very likely has to do with the fact that, as previously mentioned, we chose to excise embedded DM halos to focus our analysis on filament properties. However, while the filament radii do not decrease much as the gas approaches the halo, it is still enough to increase the central density, rising by a factor of a $\sim 5$ (dark solid line and symbols in the bottom panel of Fig. \ref{fig:ppp}). Note that such a behaviour is not specific to the gas as the DM central density (green solid line in the bottom panel of Fig. \ref{fig:ppp}) undergoes a similar change with distance to the galaxy, which is consistent with an interpretation in terms of mild gravitational focusing but also of the progressive draining of the filament core draining by the halo, as previously mentioned. Finally we want to emphasize that the {\sc nut} galaxy, essentially connected to one (two if counting each direction as an individual object) filament(s) could be somewhat of a peculiar case. In Mpc scale filaments \cite{GalarragaEspinosa2020} using the {\sc Illustris-TNG} \citep{Nelson2019} found filament properties to be dependent on filament environment, with longer filaments typically (and therefore more distant from dense structures) hosting colder gas.  Once again, further high resolution work on a much larger sample of filament/galaxies is required to  investigate the influence of connectivity and/or halo mass on the results for filaments on kpc scales.

\subsection{The impact of stellar feedback on filaments}

Stellar feedback has a profound impact on the region surrounding the galaxy and filament. Given enough time, the superbubbles it generates extend most of the way up the filament, as can be seen in the central and bottom right panels of Fig.~\ref{fig:merger}.  These galactic winds inject vorticity on large scales, and as such, this physical quantity is no longer confined to the filamentary gas. Note that, in spite of this, larger scale cosmic web filaments (i.e. larger than the superbubble) could still have well defined vorticity quadrants. More importantly, vorticity in the dark matter filament counterpart (bottom left panel of Fig.~\ref{fig:merger}) remains by-and-large unaffected, to the point that we do not deem it necessary to plot it on Fig. \ref{fig:ppp} for the run with feedback.  

The CGM/IGM gas is also strongly heated by this stellar feedback, and so the temperature signature of the accretion shock onto the filament is lost as well, as the middle right panel of Fig~\ref{fig:merger} demonstrates. Once again, this signature survives in the velocity dispersion of the DM component (middle left panel of Fig~\ref{fig:merger}). Despite such significant perturbations, the presence of a DM filament potential well coupled to the relatively high density of the gas ensures that the cooling time within the filament remains short. As a result, the filament is still visible as a cold stream cutting through the hot superbubble in the middle right panel of Fig.~\ref{fig:merger}.
 
 Due to these consequent perturbations induced by the stellar feedback, we cannot use either the temperature or vorticity to define the gas filament truncation radii in the feedback run.  It should also be noted that our assumption of isothermality of the filaments becomes less valid than in the no stellar feedback case as stronger temperature gradients develop 
 between core and outer envelope. To be more specific, in the no feedback case, the temperature varies between core and truncation radius by about a factor of two, but in the feedback case in can reach an order of magnitude.  
However, most of this gradient is localised in the outer parts of the filament, so that the central region retains a significantly large isothermal core. This can be understood 
by performing the following simple calculation.
Neglecting the presence of the wall, we may integrate both the gas and DM density profiles (from Eq. \ref{eq:plummerprof}) to obtain the filament mass per unit length, $\mu$, and its half-mass radius:
\begin{equation}
   \label{eqn:mpul}
   \begin{split}
    \mu(r)= \frac{\rho_0 \, \pi r^2}{1+(r/r_0)^2}\quad \text{and} \quad
 r_{1/2}=r_0.
\end{split}
\end{equation}
The fact that the (small) core radius contains half of the mass 
makes the filamentary material relatively impervious to the stellar feedback/filament interaction: provided the core is shielded from it, there can only be a minor change in the amount of gas mass the filament carries. It has been suggested in the literature that Kelvin-Helmholtz instabilities could be triggered at the interface between cold filament gas and the feedback powered, hot, galactic wind \cite[e.g.][and subsequent work]{Mandelker2016}. These will depend non trivially on redshift and the distance of a filament segment to the central galaxy, so it is quite difficult to define a unique characteristic timescale, $t_{\rm KH}$. Nevertheless, writing $t_{\rm KH}(r) =(r/v_w) \sqrt{\rho(r)/\rho_w}$ where $v_w$ is the relative velocity between the wind and the gas filament and $\rho_w$ the density of the wind, we can see that given the steepness of the filament density profile we measure, $t_{\rm KH}$ becomes larger as the perturbation progresses deeper in the filament. This means that the timescale is ultimately set by $t_{\rm KH}(r_0)$. Plugging in typical numbers for our feedback run at $z=4$, i.e. $r_0 \sim 5$ kpc, $\rho(r_0) \sim 3 \times 10^{-27}$g~cm$^{-3}$, $v_w \sim 100$~km/s, and $\rho_w \sim 3 \times 10^{-28}$g~cm$^{-3}$, we thus get $t_{\rm KH} \sim 20$~Myr which is about an order of magnitude shorter than the infall time from the virial  radius of the embedded halo. The conclusion is thus that our gas filaments should not survive the interaction. 

Notwithstanding that this does not happen in our simulations, which might admittedly be of too low a resolution to capture the instability properly, the calculation ignores both the importance of radiative cooling within the filament which might confine the perturbations at the surface \cite{Vietri1997}, and the important fact that, as we have previously discussed, gas filaments are {\it not} self-gravitating but are located within a dominant DM filament potential well. Because of this, it is unclear as to whether Kelvin-Helmholtz instabilities can impart to the gas a radial velocity (as in perpendicular to the filament axis) larger than the escape velocity necessary to climb out of this potential well. Should they not, they would simply render the gas flow within the filament turbulent without affecting the filamentary nature of gas accretion onto halos.

\begin{figure}
\begin{tikzpicture}
\begin{scope}[xshift=-1.5cm]
    \node[anchor=south west,inner sep=0] (image) at (0,0) {\includegraphics[width=0.9\columnwidth]{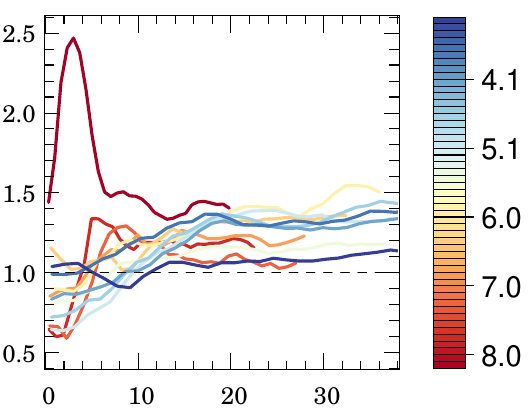}};   
\end{scope}
\node [anchor=west,rotate=90] (note) at (-1.8,2.4) {\Large $\rho_{\rm fb}/\rho_{\rm nofb}$};
\node [anchor=west] (note) at (1.2,-0.3) {\Large Radius (kpc)};
\node [anchor=west] (note) at (4.65,0.3) {\huge z};
\end{tikzpicture}
  \caption{Ratio of the median gas density profiles in the feedback and no feedback runs, $\rho_{\rm fb}/\rho_{\rm nofb}$, as a function of distance to the filament center. Curves of different colours represent different redshifts, as indicated on the figure. Very early in the simulation ($z=8$) feedback enhances the density of gas in the filament. However, at almost all other redshifts, the reverse happens: the filament is depleted of gas in the feedback run as compared to the no-feedback run. The amplitude of the effect is {\it not} monotonic with redshift.
    }
    \label{fig:reduction}

\end{figure}
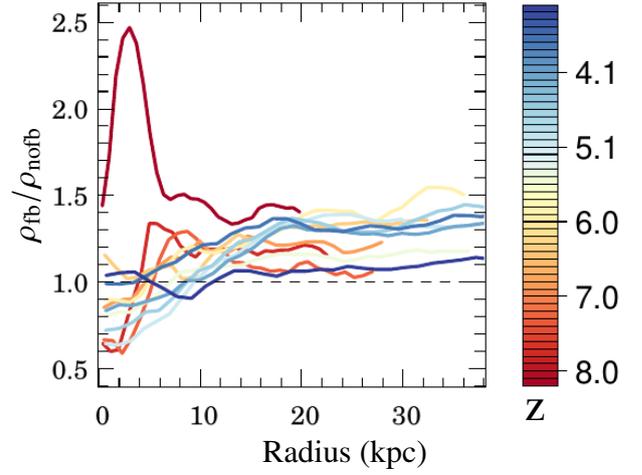

In the feedback case, we measure that the core radius of the gas filament evolves with redshift as $r_0 \propto (1+z)^{\rofb\pm\rofberr}$, i.e. with a scaling very similar to the no feedback run (see Fig \ref{fig:radius}).
Nevertheless, given the importance of the stellar feedback perturbations, one expects gas accretion onto the filament to be reduced in their presence.  To quantify this effect, we plot the ratio of median feedback to no feedback gas density profiles along the filament as a function of redshift in Fig.~\ref{fig:reduction}.
From the figure, one can see that while the size of core radius is not significantly affected by feedback, the central density is, to a larger extent.  At $z=7$ a 40\% reduction is measured, though this falls to 20\% at $z=3.6$, at which point the feedback ceases to have an effect on the filament core.  We emphasize that contrary to the growth of the core/truncation radii, the impact of feedback does not scale monotonically with redshift, as it depends both on the global properties of the IGM/filament {\it and} the star formation history of the galaxy which drives the feedback. Indeed, as shown on Fig.~\ref{fig:reduction}, at early times ($z\sim 8$) the filament core density  is even {\it enhanced} by the action of feedback.  It is possible that some of this extra gas will be entrained in the filament, but another possibility is that it will act as a shield from fresh feedback at later times.  In a future paper, we plan to use tracer particles developed in \cite{Cadiou2018} to distinguish between these two situations.  Outside the filament, the density is seen to be enhanced in the simulation with feedback, which is somehow expected from mass conservation of the filamentary gas and the presence of the extra material brought by the galactic winds.

It should be noted that the stellar feedback implemented in our simulation is the supernova prescription of \cite{KimmNUT2015}, which ensures that the correct energy/momentum is given to the gas irrespective of whether the Taylor-Sedov phase of the supernova is spatially resolved.  As such if the filaments are not destroyed by this supernova feedback then they are unlikely to be destroyed by any 'realistic' supernova feedback. Yet, other types of stellar feedback are also present which could alter filament properties, whether by direct action of the feedback on the filaments or through suppression of star formation and thereby the supernova feedback \citep[e.g. resonant scattering of Lyman alpha photons in high redshift dwarf galaxies][]{Kimm2018}. There is, of course, photo-heating due to ionising radiation which can induce an important gas density depletion especially in filaments connecting low mass halos \citep[see][for detail]{Katz2019}. We believe that this effect is, by-and-large, captured by the UV background model implementation present in both the stellar feedback and no-feedback runs. However, another mode of stellar feedback which we do not account for, might be more effective at filament disruption as it is less confined to the galaxy: cosmic rays \citep[see e.g.][]{Pfrommer2017}, . Finally, for filaments connecting halos of higher mass, \cite{DuboisAGNcluster2013} showed that AGN are also very effective at disrupting filaments, and can even destroy their cores.

\section{Conclusions} \label{sec:Conclusions}

Theory suggests \citep[e.g.][]{Keres05,Dekel06,Pichon11} that filaments play an extremely important role in the evolution of galaxies at high redshift. However, their basic characteristics are, as yet, not completely understood, and they are extremely hard to detect observationally. We used a suite of high resolution cosmological zoom-in simulations, progressively including more of the relevant physics, to place constraints on the physical properties of such a filament, from large (Mpc) scales to the point where it connects to the virial sphere of the central galaxy. 
Our main findings are as follows:
\begin{itemize}
    \item The filament in both DM and gas simulations can be described fairly accurately by a universal density profile $\rho=\frac{\rho_o}{(1+(r/r_0)^2)^2}$ corresponding to a cylinder in isothermal equilibrium 
    \item the filament core radius evolves for the gas grows as $r_0 \propto (1+z)^{\rono\pm\ronoerr}$, with the DM filament core evolving as $r_0 \propto (1+z)^{\ronoDM\pm\ronoDMerr}$. This evolution of $r_0$ for the gas closely tracks that of the size of the galaxy ($0.2 r_{\rm vir}$).
    \item The filament has a second characteristic radius, the truncation radius, which is detectable (at least in simulations) in the temperature/velocity dispersion or vorticity fields.  This radius scales as $r_{\rm tr}\propto(1+z)^{\coDMno\pm\coDMnoerr}$ for DM and $r_{\rm tr}\propto(1+z)^{\cono\pm\conoerr}$ for the gas.  The DM truncation radius closely matches the virial radius of the galaxy. The gas truncation radius is generally thinner at early times. 
    \item The filament properties are mildly affected by stellar feedback from the central galaxy. The core radius of the gas filament hardly changes, but its central density is generally reduced by $\sim$20-30 percent, but this does not happen monotonically with redshift. The DM filament properties hardly undergo any change.
\end{itemize}

Our simulations also establish that filaments need to be resolved with a minimum of $\sim 2$ kpc for a Milky Way sized halo in order to capture the filament properties.  This  might have important consequences for the angular momentum content of the gas transported to the galaxy. Still higher resolution will be required to capture the filaments around dwarf galaxies, though these are far more vulnerable to photoionisation and so probably do not need be resolved in detail beyond $z=6$. 
While the mass brought by the inflowing filament gas is affected to a level of $\sim$20-30 percent as a result of stellar feedback from the central galaxy, a further reduction is likely to occur as the filament enters the virial radius. We plan to tackle this issue using tracer particles in the near future.  The interaction of the filament with galactic winds and the virialised halo hot atmosphere will also be a function of the halo mass, and therefore our results need to be extended to a larger sample.

Indeed, our analysis was performed on the filament feeding one galaxy at high resolution. We thus plan to apply the techniques developed in this paper to the {\sc New-Horizon} simulation \citep[][Dubois et al, in prep]{park2019}, a cosmological zoom of the Horizon-AGN \citep{Dubois2014} which will have tens of galaxies of a similar stellar mass
to the one we studied in this paper, along with several more massive objects. {\sc New-Horizon} also features AGN feedback and has reached $z=0.25$, which also allows to comprehensively extend the redshift range of the analysis. Such a simulation will thus permit the extraction of a large sample of filaments from which to derive statistically meaningful quantities.   
\\

\section*{Acknowledgements}

We acknowledge useful discussions with Christophe Pichon and Harley Katz at various stages of this paper.
This work used the DiRAC Data Centric system at Durham University, operated by the Institute for Computational Cosmology on behalf of the STFC DiRAC HPC Facility (www.dirac.ac.uk. This equipment was funded by a BIS National E-infrastructure capital grant ST/K00042X/1, STFC capital grant ST/K00087X/1, DiRAC Operations grant ST/K003267/1 and Durham University. DiRAC is part of the National E-Infrastructure. JD and AS are supported by Adrian Beecroft and the STFC, and CL by a Beecroft Fellowship. 
We  warmly thank S.~Rouberol for running  the  horizon cluster on which part of the simulation was  post-processed. We thank D.~Munro for freely distributing his {\sc Yorick} programming language and opengl interface (available at \url{http://yorick.sourceforge.net/}).

\section*{Data Availability Statement}
The data underlying this article will be shared on reasonable request to the corresponding author.



\bibliographystyle{mnras}
\bibliography{biblio.bib}

\begin{thebibliography}{}
\makeatletter
\relax
\def\mn@urlcharsother{\let\do\@makeother \do\$\do\&\do\#\do\^\do\_\do\%\do\~}
\def\mn@doi{\begingroup\mn@urlcharsother \@ifnextchar [ {\mn@doi@}
  {\mn@doi@[]}}
\def\mn@doi@[#1]#2{\def\@tempa{#1}\ifx\@tempa\@empty \href
  {http://dx.doi.org/#2} {doi:#2}\else \href {http://dx.doi.org/#2} {#1}\fi
  \endgroup}
\def\mn@eprint#1#2{\mn@eprint@#1:#2::\@nil}
\def\mn@eprint@arXiv#1{\href {http://arxiv.org/abs/#1} {{\tt arXiv:#1}}}
\def\mn@eprint@dblp#1{\href {http://dblp.uni-trier.de/rec/bibtex/#1.xml}
  {dblp:#1}}
\def\mn@eprint@#1:#2:#3:#4\@nil{\def\@tempa {#1}\def\@tempb {#2}\def\@tempc
  {#3}\ifx \@tempc \@empty \let \@tempc \@tempb \let \@tempb \@tempa \fi \ifx
  \@tempb \@empty \def\@tempb {arXiv}\fi \@ifundefined
  {mn@eprint@\@tempb}{\@tempb:\@tempc}{\expandafter \expandafter \csname
  mn@eprint@\@tempb\endcsname \expandafter{\@tempc}}}

\bibitem[\protect\citeauthoryear{{Arag{\'o}n-Calvo}, {van de Weygaert}, {Jones}
   \& {van der Hulst}}{{Arag{\'o}n-Calvo} et~al.}{2007}]{AragonCalvo2007}
{Arag{\'o}n-Calvo} M.~A.,  {van de Weygaert} R.,  {Jones} B.~J.~T.,   {van der
  Hulst} J.~M.,  2007, \mn@doi [\apjl] {10.1086/511633}, \href
  {http://adsabs.harvard.edu/abs/2007ApJ...655L...5A} {655, L5}

\bibitem[\protect\citeauthoryear{{Arag{\'o}n-Calvo}, {van de Weygaert}  \&
  {Jones}}{{Arag{\'o}n-Calvo} et~al.}{2010}]{AragonCalvo2010}
{Arag{\'o}n-Calvo} M.~A.,  {van de Weygaert} R.,   {Jones} B.~J.~T.,  2010,
  \mn@doi [\mnras] {10.1111/j.1365-2966.2010.17263.x}, \href
  {http://adsabs.harvard.edu/abs/2010MNRAS.408.2163A} {408, 2163}

\bibitem[\protect\citeauthoryear{{Berlok} \& {Pfrommer}}{{Berlok} \&
  {Pfrommer}}{2019}]{Berlok2019}
{Berlok} T.,  {Pfrommer} C.,  2019, \mn@doi [\mnras] {10.1093/mnras/stz379},
  \href {https://ui.adsabs.harvard.edu/abs/2019MNRAS.485..908B} {485, 908}

\bibitem[\protect\citeauthoryear{{Bigiel} et~al.,}{{Bigiel}
  et~al.}{2011}]{Bigiel2011}
{Bigiel} F.,  et~al., 2011, \mn@doi [\apjl] {10.1088/2041-8205/730/2/L13},
  \href {http://adsabs.harvard.edu/abs/2011ApJ...730L..13B} {730, L13}

\bibitem[\protect\citeauthoryear{{Birnboim} \& {Dekel}}{{Birnboim} \&
  {Dekel}}{2003}]{Birnboim2003}
{Birnboim} Y.,  {Dekel} A.,  2003, \mn@doi [\mnras]
  {10.1046/j.1365-8711.2003.06955.x}, \href
  {https://ui.adsabs.harvard.edu/\#abs/2003MNRAS.345..349B} {345, 349}

\bibitem[\protect\citeauthoryear{{Bond}, {Kofman}  \& {Pogosyan}}{{Bond}
  et~al.}{1996}]{Bond96}
{Bond} J.~R.,  {Kofman} L.,   {Pogosyan} D.,  1996, \mn@doi [\nat]
  {10.1038/380603a0}, \href {http://adsabs.harvard.edu/abs/1996Natur.380..603B}
  {380, 603}

\bibitem[\protect\citeauthoryear{{Cadiou}, {Dubois}  \& {Pichon}}{{Cadiou}
  et~al.}{2019}]{Cadiou2018}
{Cadiou} C.,  {Dubois} Y.,   {Pichon} C.,  2019, \mn@doi [\aap]
  {10.1051/0004-6361/201834496}, \href
  {https://ui.adsabs.harvard.edu/abs/2019A&A...621A..96C} {621, A96}

\bibitem[\protect\citeauthoryear{{Cautun}, {van de Weygaert}, {Jones}  \&
  {Frenk}}{{Cautun} et~al.}{2014}]{Cautun2014}
{Cautun} M.,  {van de Weygaert} R.,  {Jones} B. J.~T.,   {Frenk} C.~S.,  2014,
  \mn@doi [\mnras] {10.1093/mnras/stu768}, \href
  {https://ui.adsabs.harvard.edu/abs/2014MNRAS.441.2923C} {441, 2923}

\bibitem[\protect\citeauthoryear{{Chabrier}}{{Chabrier}}{2003}]{Chabrier2003}
{Chabrier} G.,  2003, \mn@doi [\pasp] {10.1086/376392}, \href
  {https://ui.adsabs.harvard.edu/abs/2003PASP..115..763C} {115, 763}

\bibitem[\protect\citeauthoryear{{Chen}, {Ho}, {Blazek}, {He}, {Mandelbaum},
  {Melchior}  \& {Singh}}{{Chen} et~al.}{2019}]{Chen2018}
{Chen} Y.-C.,  {Ho} S.,  {Blazek} J.,  {He} S.,  {Mandelbaum} R.,  {Melchior}
  P.,   {Singh} S.,  2019, \mn@doi [\mnras] {10.1093/mnras/stz539}, \href
  {https://ui.adsabs.harvard.edu/abs/2019MNRAS.485.2492C} {485, 2492}

\bibitem[\protect\citeauthoryear{{Codis}, {Pichon}, {Devriendt}, {Slyz},
  {Pogosyan}, {Dubois}  \& {Sousbie}}{{Codis} et~al.}{2012}]{Codis12}
{Codis} S.,  {Pichon} C.,  {Devriendt} J.,  {Slyz} A.,  {Pogosyan} D.,
  {Dubois} Y.,   {Sousbie} T.,  2012, \mn@doi [Monthly Notices of the Royal
  Astronomical Society] {10.1111/j.1365-2966.2012.21636.x}, 427, 3320

\bibitem[\protect\citeauthoryear{{Colberg}, {Krughoff}  \&
  {Connolly}}{{Colberg} et~al.}{2005}]{Colberg2005}
{Colberg} J.~M.,  {Krughoff} K.~S.,   {Connolly} A.~J.,  2005, \mn@doi [\mnras]
  {10.1111/j.1365-2966.2005.08897.x}, \href
  {https://ui.adsabs.harvard.edu/\#abs/2005MNRAS.359..272C} {359, 272}

\bibitem[\protect\citeauthoryear{{Cornuault}, {Lehnert}, {Boulanger}  \&
  {Guillard}}{{Cornuault} et~al.}{2018}]{Cornuault2018}
{Cornuault} N.,  {Lehnert} M.~D.,  {Boulanger} F.,   {Guillard} P.,  2018,
  \mn@doi [\aap] {10.1051/0004-6361/201629229}, \href
  {http://adsabs.harvard.edu/abs/2018A%26A...610A..75C} {610, A75}

\bibitem[\protect\citeauthoryear{{Croton} et~al.,}{{Croton}
  et~al.}{2006}]{Croton2006}
{Croton} D.~J.,  et~al., 2006, \mn@doi [\mnras]
  {10.1111/j.1365-2966.2005.09675.x}, \href
  {https://ui.adsabs.harvard.edu/abs/2006MNRAS.365...11C} {365, 11}

\bibitem[\protect\citeauthoryear{{Danovich}, {Dekel}, {Hahn}  \&
  {Teyssier}}{{Danovich} et~al.}{2012}]{Danovich2012}
{Danovich} M.,  {Dekel} A.,  {Hahn} O.,   {Teyssier} R.,  2012, \mn@doi
  [\mnras] {10.1111/j.1365-2966.2012.20751.x}, \href
  {https://ui.adsabs.harvard.edu/\#abs/2012MNRAS.422.1732D} {422, 1732}

\bibitem[\protect\citeauthoryear{{Dav{\'e}}, {Angl{\'e}s-Alc{\'a}zar},
  {Narayanan}, {Li}, {Rafieferantsoa}  \& {Appleby}}{{Dav{\'e}}
  et~al.}{2019}]{DaveSimba2019}
{Dav{\'e}} R.,  {Angl{\'e}s-Alc{\'a}zar} D.,  {Narayanan} D.,  {Li} Q.,
  {Rafieferantsoa} M.~H.,   {Appleby} S.,  2019, \mn@doi [\mnras]
  {10.1093/mnras/stz937}, \href
  {https://ui.adsabs.harvard.edu/abs/2019MNRAS.486.2827D} {486, 2827}

\bibitem[\protect\citeauthoryear{{Davis}, {Huchra}, {Latham}  \&
  {Tonry}}{{Davis} et~al.}{1982}]{Davis1982}
{Davis} M.,  {Huchra} J.,  {Latham} D.~W.,   {Tonry} J.,  1982, \mn@doi [\apj]
  {10.1086/159646}, \href {http://adsabs.harvard.edu/abs/1982ApJ...253..423D}
  {253, 423}

\bibitem[\protect\citeauthoryear{{Dekel} \& {Birnboim}}{{Dekel} \&
  {Birnboim}}{2006}]{Dekel06}
{Dekel} A.,  {Birnboim} Y.,  2006, \mn@doi [\mnras]
  {10.1111/j.1365-2966.2006.10145.x}, \href
  {http://adsabs.harvard.edu/abs/2006MNRAS.368....2D} {368, 2}

\bibitem[\protect\citeauthoryear{{Dekel} et~al.,}{{Dekel}
  et~al.}{2009}]{Dekel2009}
{Dekel} A.,  et~al., 2009, \mn@doi [\nat] {10.1038/nature07648}, \href
  {https://ui.adsabs.harvard.edu/\#abs/2009Natur.457..451D} {457, 451}

\bibitem[\protect\citeauthoryear{{Diemer}, {Mansfield}, {Kravtsov}  \&
  {More}}{{Diemer} et~al.}{2017}]{Diemer2017}
{Diemer} B.,  {Mansfield} P.,  {Kravtsov} A.~V.,   {More} S.,  2017, \mn@doi
  [\apj] {10.3847/1538-4357/aa79ab}, \href
  {https://ui.adsabs.harvard.edu/abs/2017ApJ...843..140D} {843, 140}

\bibitem[\protect\citeauthoryear{Dijkstra}{Dijkstra}{1959}]{Dijkstra1959}
Dijkstra E.~W.,  1959, \mn@doi [Numerische Mathematik] {10.1007/BF01386390}, 1,
  269

\bibitem[\protect\citeauthoryear{{Dolag}, {Meneghetti}, {Moscardini}, {Rasia}
  \& {Bonaldi}}{{Dolag} et~al.}{2006}]{Dolag2006}
{Dolag} K.,  {Meneghetti} M.,  {Moscardini} L.,  {Rasia} E.,   {Bonaldi} A.,
  2006, \mn@doi [\mnras] {10.1111/j.1365-2966.2006.10511.x}, \href
  {https://ui.adsabs.harvard.edu/\#abs/2006MNRAS.370..656D} {370, 656}

\bibitem[\protect\citeauthoryear{{Dubois} \& {Teyssier}}{{Dubois} \&
  {Teyssier}}{2008}]{Dubois2008}
{Dubois} Y.,  {Teyssier} R.,  2008, \mn@doi [\aap]
  {10.1051/0004-6361:20078326}, \href
  {https://ui.adsabs.harvard.edu/abs/2008A&A...477...79D} {477, 79}

\bibitem[\protect\citeauthoryear{{Dubois}, {Pichon}, {Devriendt}, {Silk},
  {Haehnelt}, {Kimm}  \& {Slyz}}{{Dubois} et~al.}{2013}]{DuboisAGNcluster2013}
{Dubois} Y.,  {Pichon} C.,  {Devriendt} J.,  {Silk} J.,  {Haehnelt} M.,  {Kimm}
  T.,   {Slyz} A.,  2013, \mn@doi [\mnras] {10.1093/mnras/sts224}, \href
  {https://ui.adsabs.harvard.edu/\#abs/2013MNRAS.428.2885D} {428, 2885}

\bibitem[\protect\citeauthoryear{{Dubois} et~al.,}{{Dubois}
  et~al.}{2014}]{Dubois2014}
{Dubois} Y.,  et~al., 2014, \mn@doi [\mnras] {10.1093/mnras/stu1227}, \href
  {http://adsabs.harvard.edu/abs/2014MNRAS.444.1453D} {444, 1453}

\bibitem[\protect\citeauthoryear{{Dunkley} et~al.,}{{Dunkley}
  et~al.}{2009}]{Dunkley09}
{Dunkley} J.,  et~al., 2009, \mn@doi [\apjs] {10.1088/0067-0049/180/2/306},
  \href {http://adsabs.harvard.edu/abs/2009ApJS..180..306D} {180, 306}

\bibitem[\protect\citeauthoryear{{Elias}, {Genel}, {Sternberg}, {Devriendt},
  {Slyz}, {Visbal}  \& {Bouch{\'e}}}{{Elias} et~al.}{2020}]{Lydia2020}
{Elias} L.~M.,  {Genel} S.,  {Sternberg} A.,  {Devriendt} J.,  {Slyz} A.,
  {Visbal} E.,   {Bouch{\'e}} N.,  2020, \mn@doi [\mnras]
  {10.1093/mnras/staa1059}, \href
  {https://ui.adsabs.harvard.edu/abs/2020MNRAS.494.5439E} {494, 5439}

\bibitem[\protect\citeauthoryear{{Freundlich}, {Jog}  \& {Combes}}{{Freundlich}
  et~al.}{2014}]{Freundlich2014}
{Freundlich} J.,  {Jog} C.~J.,   {Combes} F.,  2014, \mn@doi [\aap]
  {10.1051/0004-6361/201323325}, \href
  {https://ui.adsabs.harvard.edu/abs/2014A&A...564A...7F} {564, A7}

\bibitem[\protect\citeauthoryear{{Gallego} et~al.,}{{Gallego}
  et~al.}{2018}]{Gallego17}
{Gallego} S.~G.,  et~al., 2018, \mn@doi [\mnras] {10.1093/mnras/sty037}, \href
  {https://ui.adsabs.harvard.edu/abs/2018MNRAS.475.3854G} {475, 3854}

\bibitem[\protect\citeauthoryear{Galárraga-Espinosa, Aghanim, Langer  \&
  Tanimura}{Galárraga-Espinosa et~al.}{2020}]{GalarragaEspinosa2020}
Galárraga-Espinosa D.,  Aghanim N.,  Langer M.,   Tanimura H.,  2020,
  Properties of gas phases around cosmic filaments at z=0 in the Illustris-TNG
  simulation (\mn@eprint {arXiv} {2010.15139})

\bibitem[\protect\citeauthoryear{{Ganeshaiah Veena}, {Cautun}, {van de
  Weygaert}, {Tempel}, {Jones}, {Rieder}  \& {Frenk}}{{Ganeshaiah Veena}
  et~al.}{2018}]{Ganesh2018}
{Ganeshaiah Veena} P.,  {Cautun} M.,  {van de Weygaert} R.,  {Tempel} E.,
  {Jones} B.~J.~T.,  {Rieder} S.,   {Frenk} C.~S.,  2018, \mn@doi [\mnras]
  {10.1093/mnras/sty2270}, \href
  {http://adsabs.harvard.edu/abs/2018MNRAS.481..414G} {481, 414}

\bibitem[\protect\citeauthoryear{{Gao} \& {Theuns}}{{Gao} \&
  {Theuns}}{2007}]{Gao2007}
{Gao} L.,  {Theuns} T.,  2007, \mn@doi [Science] {10.1126/science.1146676},
  \href {https://ui.adsabs.harvard.edu/abs/2007Sci...317.1527G} {317, 1527}

\bibitem[\protect\citeauthoryear{{Gardner} et~al.,}{{Gardner}
  et~al.}{2006}]{Gardner2006}
{Gardner} J.~P.,  et~al., 2006, \mn@doi [\ssr] {10.1007/s11214-006-8315-7},
  \href {https://ui.adsabs.harvard.edu/abs/2006SSRv..123..485G} {123, 485}

\bibitem[\protect\citeauthoryear{{Geller} \& {Huchra}}{{Geller} \&
  {Huchra}}{1989}]{Geller1989}
{Geller} M.~J.,  {Huchra} J.~P.,  1989, \mn@doi [Science]
  {10.1126/science.246.4932.897}, \href
  {http://adsabs.harvard.edu/abs/1989Sci...246..897G} {246, 897}

\bibitem[\protect\citeauthoryear{{Gheller}, {Vazza}, {Favre}  \&
  {Br{\"u}ggen}}{{Gheller} et~al.}{2015}]{Gheller2015}
{Gheller} C.,  {Vazza} F.,  {Favre} J.,   {Br{\"u}ggen} M.,  2015, \mn@doi
  [\mnras] {10.1093/mnras/stv1646}, \href
  {https://ui.adsabs.harvard.edu/abs/2015MNRAS.453.1164G} {453, 1164}

\bibitem[\protect\citeauthoryear{{Gheller}, {Vazza}, {Br{\"u}ggen}, {Alpaslan},
  {Holwerda}, {Hopkins}  \& {Liske}}{{Gheller} et~al.}{2016}]{Gheller2016}
{Gheller} C.,  {Vazza} F.,  {Br{\"u}ggen} M.,  {Alpaslan} M.,  {Holwerda}
  B.~W.,  {Hopkins} A.~M.,   {Liske} J.,  2016, \mn@doi [\mnras]
  {10.1093/mnras/stw1595}, \href
  {https://ui.adsabs.harvard.edu/abs/2016MNRAS.462..448G} {462, 448}

\bibitem[\protect\citeauthoryear{{Giavalisco} et~al.,}{{Giavalisco}
  et~al.}{2011}]{Giavalisco11}
{Giavalisco} M.,  et~al., 2011, \mn@doi [\apj] {10.1088/0004-637X/743/1/95},
  \href {http://adsabs.harvard.edu/abs/2011ApJ...743...95G} {743, 95}

\bibitem[\protect\citeauthoryear{{Grogin} et~al.,}{{Grogin}
  et~al.}{2011}]{Grogin2011}
{Grogin} N.~A.,  et~al., 2011, \mn@doi [\apjs] {10.1088/0067-0049/197/2/35},
  \href {https://ui.adsabs.harvard.edu/abs/2011ApJS..197...35G} {197, 35}

\bibitem[\protect\citeauthoryear{Hartwig, Bromm  \& Loeb}{Hartwig
  et~al.}{2018}]{Hartwig_2018}
Hartwig T.,  Bromm V.,   Loeb A.,  2018, \mn@doi [Monthly Notices of the Royal
  Astronomical Society] {10.1093/mnras/sty1576}, 479, 2202–2213

\bibitem[\protect\citeauthoryear{{Ho} \& {Martin}}{{Ho} \&
  {Martin}}{2019}]{Ho2019}
{Ho} S.~H.,  {Martin} C.~L.,  2019, arXiv e-prints, \href
  {https://ui.adsabs.harvard.edu/\#abs/2019arXiv190111182H} {p.
  arXiv:1901.11182}

\bibitem[\protect\citeauthoryear{{Japelj} et~al.,}{{Japelj}
  et~al.}{2019}]{Japelj2019}
{Japelj} J.,  et~al., 2019, \mn@doi [\aap] {10.1051/0004-6361/201936048}, \href
  {https://ui.adsabs.harvard.edu/abs/2019A&A...632A..94J} {632, A94}

\bibitem[\protect\citeauthoryear{{Jeans}}{{Jeans}}{1915}]{Jeans1915}
{Jeans} J.~H.,  1915, \mn@doi [\mnras] {10.1093/mnras/76.2.70}, \href
  {http://adsabs.harvard.edu/abs/1915MNRAS..76...70J} {76, 70}

\bibitem[\protect\citeauthoryear{{Kacprzak}, {Churchill}  \&
  {Nielsen}}{{Kacprzak} et~al.}{2012}]{Kaz12}
{Kacprzak} G.~G.,  {Churchill} C.~W.,   {Nielsen} N.~M.,  2012, \mn@doi [\apjl]
  {10.1088/2041-8205/760/1/L7}, \href
  {http://adsabs.harvard.edu/abs/2012ApJ...760L...7K} {760, L7}

\bibitem[\protect\citeauthoryear{{Katz}}{{Katz}}{1992}]{Katzovercooling1992}
{Katz} N.,  1992, \mn@doi [\apj] {10.1086/171366}, \href
  {http://adsabs.harvard.edu/abs/1992ApJ...391..502K} {391, 502}

\bibitem[\protect\citeauthoryear{Katz, Keres, Dave, Weinberg, Rosenberg  \&
  Putman}{Katz et~al.}{2003}]{Katz03}
Katz N.,  Keres D.,  Dave R.,  Weinberg D.~H.,  Rosenberg J.,   Putman M.,
  2003, Astrophysics and Space Science Library, 281, 185

\bibitem[\protect\citeauthoryear{{Katz} et~al.,}{{Katz}
  et~al.}{2019}]{Katz2019}
{Katz} H.,  et~al., 2019, arXiv e-prints, \href
  {https://ui.adsabs.harvard.edu/abs/2019arXiv190511414K} {p. arXiv:1905.11414}

\bibitem[\protect\citeauthoryear{{Kennicutt}}{{Kennicutt}}{1998}]{Kennicutt1998}
{Kennicutt} Robert~C. J.,  1998, \mn@doi [\apj] {10.1086/305588}, \href
  {https://ui.adsabs.harvard.edu/abs/1998ApJ...498..541K} {498, 541}

\bibitem[\protect\citeauthoryear{{Kere{\v s}}, {Katz}, {Weinberg}  \&
  {Dav{\'e}}}{{Kere{\v s}} et~al.}{2005}]{Keres05}
{Kere{\v s}} D.,  {Katz} N.,  {Weinberg} D.~H.,   {Dav{\'e}} R.,  2005, \mn@doi
  [\mnras] {10.1111/j.1365-2966.2005.09451.x}, \href
  {http://adsabs.harvard.edu/abs/2005MNRAS.363....2K} {363, 2}

\bibitem[\protect\citeauthoryear{{Khandai}, {Di Matteo}, {Croft}, {Wilkins},
  {Feng}, {Tucker}, {DeGraf}  \& {Liu}}{{Khandai} et~al.}{2015}]{Khandai2015}
{Khandai} N.,  {Di Matteo} T.,  {Croft} R.,  {Wilkins} S.,  {Feng} Y.,
  {Tucker} E.,  {DeGraf} C.,   {Liu} M.-S.,  2015, \mn@doi [\mnras]
  {10.1093/mnras/stv627}, \href
  {https://ui.adsabs.harvard.edu/abs/2015MNRAS.450.1349K} {450, 1349}

\bibitem[\protect\citeauthoryear{{Kikuta} et~al.,}{{Kikuta}
  et~al.}{2019}]{Kikuta2019}
{Kikuta} S.,  et~al., 2019, \mn@doi [\pasj] {10.1093/pasj/psz055}, \href
  {https://ui.adsabs.harvard.edu/abs/2019PASJ...71L...2K} {71, L2}

\bibitem[\protect\citeauthoryear{{Kimm}, {Devriendt}, {Slyz}, {Pichon},
  {Kassin}  \& {Dubois}}{{Kimm} et~al.}{2011}]{Kimm2011}
{Kimm} T.,  {Devriendt} J.,  {Slyz} A.,  {Pichon} C.,  {Kassin} S.~A.,
  {Dubois} Y.,  2011, preprint, \href
  {http://adsabs.harvard.edu/abs/2011arXiv1106.0538K} {} (\mn@eprint {arXiv}
  {1106.0538})

\bibitem[\protect\citeauthoryear{{Kimm}, {Cen}, {Devriendt}, {Dubois}  \&
  {Slyz}}{{Kimm} et~al.}{2015}]{KimmNUT2015}
{Kimm} T.,  {Cen} R.,  {Devriendt} J.,  {Dubois} Y.,   {Slyz} A.,  2015,
  \mn@doi [\mnras] {10.1093/mnras/stv1211}, \href
  {http://adsabs.harvard.edu/abs/2015MNRAS.451.2900K} {451, 2900}

\bibitem[\protect\citeauthoryear{{Kimm}, {Haehnelt}, {Blaizot}, {Katz},
  {Michel-Dansac}, {Garel}, {Rosdahl}  \& {Teyssier}}{{Kimm}
  et~al.}{2018}]{Kimm2018}
{Kimm} T.,  {Haehnelt} M.,  {Blaizot} J.,  {Katz} H.,  {Michel-Dansac} L.,
  {Garel} T.,  {Rosdahl} J.,   {Teyssier} R.,  2018, \mn@doi [\mnras]
  {10.1093/mnras/sty126}, \href
  {https://ui.adsabs.harvard.edu/abs/2018MNRAS.475.4617K} {475, 4617}

\bibitem[\protect\citeauthoryear{{Klar} \& {M{\"u}cket}}{{Klar} \&
  {M{\"u}cket}}{2012}]{Klar2012}
{Klar} J.~S.,  {M{\"u}cket} J.~P.,  2012, \mn@doi [\mnras]
  {10.1111/j.1365-2966.2012.20877.x}, \href
  {https://ui.adsabs.harvard.edu/\#abs/2012MNRAS.423..304K} {423, 304}

\bibitem[\protect\citeauthoryear{{Koekemoer} et~al.,}{{Koekemoer}
  et~al.}{2011}]{Kroekemoer2011}
{Koekemoer} A.~M.,  et~al., 2011, \mn@doi [\apjs] {10.1088/0067-0049/197/2/36},
  \href {https://ui.adsabs.harvard.edu/abs/2011ApJS..197...36K} {197, 36}

\bibitem[\protect\citeauthoryear{Kooistra, Silva, Zaroubi, Verheijen, Tempel
  \& Hess}{Kooistra et~al.}{2019}]{Kooistra_2019}
Kooistra R.,  Silva M.~B.,  Zaroubi S.,  Verheijen M. A.~W.,  Tempel E.,   Hess
  K.~M.,  2019, \mn@doi [Monthly Notices of the Royal Astronomical Society]
  {10.1093/mnras/stz2677}, 490, 1415–1424

\bibitem[\protect\citeauthoryear{{Kraljic}, {Dave}  \& {Pichon}}{{Kraljic}
  et~al.}{2019}]{kraljic19}
{Kraljic} K.,  {Dave} R.,   {Pichon} C.,  2019, arXiv e-prints, \href
  {https://ui.adsabs.harvard.edu/abs/2019arXiv190601623K} {p. arXiv:1906.01623}

\bibitem[\protect\citeauthoryear{{Krolewski}, {Ho}, {Chen}, {Chan}, {Tenneti},
  {Bizyaev}  \& {Kraljic}}{{Krolewski} et~al.}{2019}]{Krolewski}
{Krolewski} A.,  {Ho} S.,  {Chen} Y.-C.,  {Chan} P.~F.,  {Tenneti} A.,
  {Bizyaev} D.,   {Kraljic} K.,  2019, arXiv e-prints, \href
  {https://ui.adsabs.harvard.edu/\#abs/2019arXiv190209797K} {p.
  arXiv:1902.09797}

\bibitem[\protect\citeauthoryear{{Laigle} et~al.,}{{Laigle}
  et~al.}{2015}]{Laigle15}
{Laigle} C.,  et~al., 2015, MNRAS, 446, 2744

\bibitem[\protect\citeauthoryear{{Latif}, {Schleicher}, {Spaans}  \&
  {Zaroubi}}{{Latif} et~al.}{2011}]{Latif2011}
{Latif} M.~A.,  {Schleicher} D. R.~G.,  {Spaans} M.,   {Zaroubi} S.,  2011,
  \mn@doi [\mnras] {10.1111/j.1745-3933.2011.01026.x}, \href
  {https://ui.adsabs.harvard.edu/abs/2011MNRAS.413L..33L} {413, L33}

\bibitem[\protect\citeauthoryear{{Lee}, {Hennawi}, {White}, {Croft}  \&
  {Ozbek}}{{Lee} et~al.}{2014}]{Lee2014}
{Lee} K.-G.,  {Hennawi} J.~F.,  {White} M.,  {Croft} R. A.~C.,   {Ozbek} M.,
  2014, \mn@doi [\apj] {10.1088/0004-637X/788/1/49}, \href
  {https://ui.adsabs.harvard.edu/abs/2014ApJ...788...49L} {788, 49}

\bibitem[\protect\citeauthoryear{{Leroy} et~al.,}{{Leroy}
  et~al.}{2013}]{Leroy2013}
{Leroy} A.~K.,  et~al., 2013, \mn@doi [\aj] {10.1088/0004-6256/146/2/19}, \href
  {http://adsabs.harvard.edu/abs/2013AJ....146...19L} {146, 19}

\bibitem[\protect\citeauthoryear{{Mandelker}, {Padnos}, {Dekel}, {Birnboim},
  {Burkert}, {Krumholz}  \& {Steinberg}}{{Mandelker}
  et~al.}{2016}]{Mandelker2016}
{Mandelker} N.,  {Padnos} D.,  {Dekel} A.,  {Birnboim} Y.,  {Burkert} A.,
  {Krumholz} M.~R.,   {Steinberg} E.,  2016, \mn@doi [\mnras]
  {10.1093/mnras/stw2267}, \href
  {http://adsabs.harvard.edu/abs/2016MNRAS.463.3921M} {463, 3921}

\bibitem[\protect\citeauthoryear{{Mandelker}, {van Dokkum}, {Brodie}, {van den
  Bosch}  \& {Ceverino}}{{Mandelker} et~al.}{2018}]{Mandelker2018}
{Mandelker} N.,  {van Dokkum} P.~G.,  {Brodie} J.~P.,  {van den Bosch} F.~C.,
  {Ceverino} D.,  2018, \mn@doi [\apj] {10.3847/1538-4357/aaca98}, \href
  {https://ui.adsabs.harvard.edu/abs/2018ApJ...861..148M} {861, 148}

\bibitem[\protect\citeauthoryear{{Mandelker}, {Nagai}, {Aung}, {Dekel},
  {Padnos}  \& {Birnboim}}{{Mandelker} et~al.}{2019}]{Mandelker2019}
{Mandelker} N.,  {Nagai} D.,  {Aung} H.,  {Dekel} A.,  {Padnos} D.,
  {Birnboim} Y.,  2019, \mn@doi [\mnras] {10.1093/mnras/stz012}, \href
  {http://adsabs.harvard.edu/abs/2019MNRAS.484.1100M} {484, 1100}

\bibitem[\protect\citeauthoryear{{Martin}, {Matuszewski}, {Morrissey}, {Neill},
  {Moore}, {Steidel}  \& {Trainor}}{{Martin} et~al.}{2016}]{Martin16}
{Martin} D.,  {Matuszewski} M.,  {Morrissey} P.,  {Neill} J.,  {Moore} A.,
  {Steidel} C.,   {Trainor} R.,  2016, The Astrophysical Journal Letters, 824,
  L5

\bibitem[\protect\citeauthoryear{{Mocz} et~al.,}{{Mocz}
  et~al.}{2019}]{Mocz2019}
{Mocz} P.,  et~al., 2019, \mn@doi [\prl] {10.1103/PhysRevLett.123.141301},
  \href {https://ui.adsabs.harvard.edu/abs/2019PhRvL.123n1301M} {123, 141301}

\bibitem[\protect\citeauthoryear{{Mu{\~n}oz-Cuartas}, {Macci{\`o}},
  {Gottl{\"o}ber}  \& {Dutton}}{{Mu{\~n}oz-Cuartas} et~al.}{2011}]{Munoz2011}
{Mu{\~n}oz-Cuartas} J.~C.,  {Macci{\`o}} A.~V.,  {Gottl{\"o}ber} S.,   {Dutton}
  A.~A.,  2011, \mn@doi [\mnras] {10.1111/j.1365-2966.2010.17704.x}, \href
  {https://ui.adsabs.harvard.edu/abs/2011MNRAS.411..584M} {411, 584}

\bibitem[\protect\citeauthoryear{{Nelson} et~al.,}{{Nelson}
  et~al.}{2018}]{Nelson2018}
{Nelson} D.,  et~al., 2018, arXiv e-prints, \href
  {https://ui.adsabs.harvard.edu/\#abs/2018arXiv181205609N} {p.
  arXiv:1812.05609}

\bibitem[\protect\citeauthoryear{{Nelson} et~al.,}{{Nelson}
  et~al.}{2019a}]{Nelson2019TNG50}
{Nelson} D.,  et~al., 2019a, arXiv e-prints, \href
  {http://adsabs.harvard.edu/abs/2019arXiv190205554N} {}

\bibitem[\protect\citeauthoryear{{Nelson} et~al.,}{{Nelson}
  et~al.}{2019b}]{Nelson2019}
{Nelson} D.,  et~al., 2019b, \mn@doi [Computational Astrophysics and Cosmology]
  {10.1186/s40668-019-0028-x}, \href
  {https://ui.adsabs.harvard.edu/abs/2019ComAC...6....2N} {6, 2}

\bibitem[\protect\citeauthoryear{{Ocvirk}, {Pichon}  \& {Teyssier}}{{Ocvirk}
  et~al.}{2008}]{Ocvirk2008}
{Ocvirk} P.,  {Pichon} C.,   {Teyssier} R.,  2008, \mn@doi [\mnras]
  {10.1111/j.1365-2966.2008.13763.x}, \href
  {http://adsabs.harvard.edu/abs/2008MNRAS.390.1326O} {390, 1326}

\bibitem[\protect\citeauthoryear{{Ocvirk} et~al.,}{{Ocvirk}
  et~al.}{2016}]{Ocvirk2016}
{Ocvirk} P.,  et~al., 2016, \mn@doi [\mnras] {10.1093/mnras/stw2036}, \href
  {https://ui.adsabs.harvard.edu/\#abs/2016MNRAS.463.1462O} {463, 1462}

\bibitem[\protect\citeauthoryear{{Ostriker}}{{Ostriker}}{1964}]{Ostriker1964}
{Ostriker} J.,  1964, \mn@doi [\apj] {10.1086/148005}, \href
  {http://adsabs.harvard.edu/abs/1964ApJ...140.1056O} {140, 1056}

\bibitem[\protect\citeauthoryear{{Padnos}, {Mandelker}, {Birnboim}, {Dekel},
  {Krumholz}  \& {Steinberg}}{{Padnos} et~al.}{2018}]{Padnos2018}
{Padnos} D.,  {Mandelker} N.,  {Birnboim} Y.,  {Dekel} A.,  {Krumholz} M.~R.,
  {Steinberg} E.,  2018, \mn@doi [\mnras] {10.1093/mnras/sty789}, \href
  {http://adsabs.harvard.edu/abs/2018MNRAS.477.3293P} {477, 3293}

\bibitem[\protect\citeauthoryear{{Pandya} et~al.,}{{Pandya}
  et~al.}{2019}]{Pandya2019}
{Pandya} V.,  et~al., 2019, \mn@doi [\mnras] {10.1093/mnras/stz2129}, \href
  {https://ui.adsabs.harvard.edu/abs/2019MNRAS.488.5580P} {488, 5580}

\bibitem[\protect\citeauthoryear{{Park} et~al.,}{{Park}
  et~al.}{2019}]{park2019}
{Park} M.-J.,  et~al., 2019, arXiv e-prints, \href
  {https://ui.adsabs.harvard.edu/abs/2019arXiv190502216P} {p. arXiv:1905.02216}

\bibitem[\protect\citeauthoryear{{Pfrommer}, {Pakmor}, {Schaal}, {Simpson}  \&
  {Springel}}{{Pfrommer} et~al.}{2017}]{Pfrommer2017}
{Pfrommer} C.,  {Pakmor} R.,  {Schaal} K.,  {Simpson} C.~M.,   {Springel} V.,
  2017, \mn@doi [\mnras] {10.1093/mnras/stw2941}, \href
  {https://ui.adsabs.harvard.edu/abs/2017MNRAS.465.4500P} {465, 4500}

\bibitem[\protect\citeauthoryear{{Pichon} \& {Bernardeau}}{{Pichon} \&
  {Bernardeau}}{1999}]{Pichon1999}
{Pichon} C.,  {Bernardeau} F.,  1999, \aap, \href
  {https://ui.adsabs.harvard.edu/abs/1999A&A...343..663P} {343, 663}

\bibitem[\protect\citeauthoryear{{Pichon}, {Pogosyan}, {Kimm}, {Slyz},
  {Devriendt}  \& {Dubois}}{{Pichon} et~al.}{2011}]{Pichon11}
{Pichon} C.,  {Pogosyan} D.,  {Kimm} T.,  {Slyz} A.,  {Devriendt} J.,
  {Dubois} Y.,  2011, MNRAS, 418, 2493–2507

\bibitem[\protect\citeauthoryear{{Pogosyan}, {Bond}, {Kofman}  \&
  {Wadsley}}{{Pogosyan} et~al.}{1998}]{Dmitry}
{Pogosyan} D.,  {Bond} J.~R.,  {Kofman} L.,   {Wadsley} J.,  1998, in {Colombi}
  S.,  {Mellier} Y.,   {Raban} B.,  eds, Wide Field Surveys in Cosmology. p.~61
  (\mn@eprint {} {astro-ph/9810072})

\bibitem[\protect\citeauthoryear{{Powell}, {Slyz}  \& {Devriendt}}{{Powell}
  et~al.}{2011}]{Powell11}
{Powell} L.~C.,  {Slyz} A.,   {Devriendt} J.,  2011, \mn@doi [\mnras]
  {10.1111/j.1365-2966.2011.18668.x}, \href
  {http://adsabs.harvard.edu/abs/2011MNRAS.414.3671P} {414, 3671}

\bibitem[\protect\citeauthoryear{{Power}, {Navarro}, {Jenkins}, {Frenk},
  {White}, {Springel}, {Stadel}  \& {Quinn}}{{Power} et~al.}{2003}]{Power2003}
{Power} C.,  {Navarro} J.~F.,  {Jenkins} A.,  {Frenk} C.~S.,  {White} S.~D.~M.,
   {Springel} V.,  {Stadel} J.,   {Quinn} T.,  2003, \mn@doi [\mnras]
  {10.1046/j.1365-8711.2003.05925.x}, \href
  {https://ui.adsabs.harvard.edu/abs/2003MNRAS.338...14P} {338, 14}

\bibitem[\protect\citeauthoryear{{Prescott}, {Martin}  \& {Dey}}{{Prescott}
  et~al.}{2015}]{Prescott2015}
{Prescott} M.~K.~M.,  {Martin} C.~L.,   {Dey} A.,  2015, \mn@doi [\apj]
  {10.1088/0004-637X/799/1/62}, \href
  {http://adsabs.harvard.edu/abs/2015ApJ...799...62P} {799, 62}

\bibitem[\protect\citeauthoryear{{Prunet}, {Pichon}, {Aubert}, {Pogosyan},
  {Teyssier}  \& {Gottloeber}}{{Prunet} et~al.}{2008}]{Prunet2008}
{Prunet} S.,  {Pichon} C.,  {Aubert} D.,  {Pogosyan} D.,  {Teyssier} R.,
  {Gottloeber} S.,  2008, \mn@doi [\apjs] {10.1086/590370}, \href
  {https://ui.adsabs.harvard.edu/abs/2008ApJS..178..179P} {178, 179}

\bibitem[\protect\citeauthoryear{Rahman et~al.,}{Rahman
  et~al.}{2012}]{Rahman2012}
Rahman N.,  et~al., 2012, \mn@doi [The Astrophysical Journal]
  {10.1088/0004-637x/745/2/183}, 745, 183

\bibitem[\protect\citeauthoryear{{Rasera} \& {Teyssier}}{{Rasera} \&
  {Teyssier}}{2006}]{Rasera2006}
{Rasera} Y.,  {Teyssier} R.,  2006, \mn@doi [\aap]
  {10.1051/0004-6361:20053116}, \href
  {https://ui.adsabs.harvard.edu/abs/2006A&A...445....1R} {445, 1}

\bibitem[\protect\citeauthoryear{{Ribaudo}, {Lehner}, {Howk}, {Werk}, {Tripp},
  {Prochaska}, {Meiring}  \& {Tumlinson}}{{Ribaudo} et~al.}{2011}]{Ribaudo11}
{Ribaudo} J.,  {Lehner} N.,  {Howk} J.~C.,  {Werk} J.~K.,  {Tripp} T.~M.,
  {Prochaska} J.~X.,  {Meiring} J.~D.,   {Tumlinson} J.,  2011, \mn@doi [\apj]
  {10.1088/0004-637X/743/2/207}, \href
  {http://adsabs.harvard.edu/abs/2011ApJ...743..207R} {743, 207}

\bibitem[\protect\citeauthoryear{{Rosdahl} \& {Blaizot}}{{Rosdahl} \&
  {Blaizot}}{2012}]{Rosdahl2012}
{Rosdahl} J.,  {Blaizot} J.,  2012, \mn@doi [\mnras]
  {10.1111/j.1365-2966.2012.20883.x}, \href
  {https://ui.adsabs.harvard.edu/\#abs/2012MNRAS.423..344R} {423, 344}

\bibitem[\protect\citeauthoryear{{Rosen} \& {Bregman}}{{Rosen} \&
  {Bregman}}{1995}]{Rosen1995}
{Rosen} A.,  {Bregman} J.~N.,  1995, \mn@doi [\apj] {10.1086/175303}, \href
  {http://adsabs.harvard.edu/abs/1995ApJ...440..634R} {440, 634}

\bibitem[\protect\citeauthoryear{{Schaap}}{{Schaap}}{2007}]{DTFE2007}
{Schaap} W.~E.,  2007, PhD thesis, Kapteyn Astronomical Institute
  <EMAIL>w\_schaap@live.nl</EMAIL>

\bibitem[\protect\citeauthoryear{{Schaap} \& {van de Weygaert}}{{Schaap} \&
  {van de Weygaert}}{2000}]{SchappetVandeWeygaert2000}
{Schaap} W.~E.,  {van de Weygaert} R.,  2000, \aap, \href
  {http://adsabs.harvard.edu/abs/2000A%26A...363L..29S} {363, L29}

\bibitem[\protect\citeauthoryear{{Schaye} et~al.,}{{Schaye}
  et~al.}{2015}]{Schaye2015}
{Schaye} J.,  et~al., 2015, \mn@doi [\mnras] {10.1093/mnras/stu2058}, \href
  {https://ui.adsabs.harvard.edu/\#abs/2015MNRAS.446..521S} {446, 521}

\bibitem[\protect\citeauthoryear{{Sousbie}}{{Sousbie}}{2011}]{Sousbie11}
{Sousbie} T.,  2011, \mn@doi [\mnras] {10.1111/j.1365-2966.2011.18394.x}, \href
  {http://adsabs.harvard.edu/abs/2011MNRAS.414..350S} {414, 350}

\bibitem[\protect\citeauthoryear{Spitzer}{Spitzer}{1978}]{Bookism1978}
Spitzer L.,  1978, Physical Processes in the Interstellar Medium.
New York, \url {https://books.google.co.uk/books?id=ytK2swEACAAJ}

\bibitem[\protect\citeauthoryear{{Stewart}, {Brooks}, {Bullock}, {Maller},
  {Diemand}, {Wadsley}  \& {Moustakas}}{{Stewart} et~al.}{2013}]{Stewart2013}
{Stewart} K.~R.,  {Brooks} A.~M.,  {Bullock} J.~S.,  {Maller} A.~H.,  {Diemand}
  J.,  {Wadsley} J.,   {Moustakas} L.~A.,  2013, \mn@doi [\apj]
  {10.1088/0004-637X/769/1/74}, \href
  {http://adsabs.harvard.edu/abs/2013ApJ...769...74S} {769, 74}

\bibitem[\protect\citeauthoryear{{Stewart} et~al.,}{{Stewart}
  et~al.}{2017}]{Stewart2017}
{Stewart} K.~R.,  et~al., 2017, \mn@doi [\apj] {10.3847/1538-4357/aa6dff},
  \href {http://adsabs.harvard.edu/abs/2017ApJ...843...47S} {843, 47}

\bibitem[\protect\citeauthoryear{{Stod{\'o}lkiewicz}}{{Stod{\'o}lkiewicz}}{1963}]{Stodolkiewicz1963}
{Stod{\'o}lkiewicz} J.~S.,  1963, \actaa, \href
  {http://adsabs.harvard.edu/abs/1963AcA....13...30S} {13, 30}

\bibitem[\protect\citeauthoryear{{Sutherland} \& {Dopita}}{{Sutherland} \&
  {Dopita}}{1993}]{toocool4school}
{Sutherland} R.~S.,  {Dopita} M.~A.,  1993, \mn@doi [\apjs] {10.1086/191823},
  \href {http://adsabs.harvard.edu/abs/1993ApJS...88..253S} {88, 253}

\bibitem[\protect\citeauthoryear{{Tempel} \& {Libeskind}}{{Tempel} \&
  {Libeskind}}{2013}]{Tempel13}
{Tempel} E.,  {Libeskind} N.~I.,  2013, \mn@doi [\apjl]
  {10.1088/2041-8205/775/2/L42}, \href
  {http://adsabs.harvard.edu/abs/2013ApJ...775L..42T} {775, L42}

\bibitem[\protect\citeauthoryear{{Teyssier}}{{Teyssier}}{2002}]{Teyssier02}
{Teyssier} R.,  2002, \mn@doi [\aap] {10.1051/0004-6361:20011817}, \href
  {http://adsabs.harvard.edu/abs/2002A%26A...385..337T} {385, 337}

\bibitem[\protect\citeauthoryear{{Tillson}, {Devriendt}, {Slyz}, {Miller}  \&
  {Pichon}}{{Tillson} et~al.}{2015}]{Tillson15}
{Tillson} H.,  {Devriendt} J.,  {Slyz} A.,  {Miller} L.,   {Pichon} C.,  2015,
  MNRAS, 449, 4363

\bibitem[\protect\citeauthoryear{{Umehata} et~al.,}{{Umehata}
  et~al.}{2019}]{Umehata2019}
{Umehata} H.,  et~al., 2019, \mn@doi [Science] {10.1126/science.aaw5949}, \href
  {https://ui.adsabs.harvard.edu/abs/2019Sci...366...97U} {366, 97}

\bibitem[\protect\citeauthoryear{{Vietri}, {Ferrara}  \& {Miniati}}{{Vietri}
  et~al.}{1997}]{Vietri1997}
{Vietri} M.,  {Ferrara} A.,   {Miniati} F.,  1997, \mn@doi [\apj]
  {10.1086/304202}, \href
  {https://ui.adsabs.harvard.edu/abs/1997ApJ...483..262V} {483, 262}

\bibitem[\protect\citeauthoryear{{Vogelsberger} et~al.,}{{Vogelsberger}
  et~al.}{2014}]{Vogelsberger2014}
{Vogelsberger} M.,  et~al., 2014, \mn@doi [\mnras] {10.1093/mnras/stu1536},
  \href {http://adsabs.harvard.edu/abs/2014MNRAS.444.1518V} {444, 1518}

\bibitem[\protect\citeauthoryear{{Woods}, {Wadsley}, {Couchman}, {Stinson}  \&
  {Shen}}{{Woods} et~al.}{2014}]{Woods14}
{Woods} R.~M.,  {Wadsley} J.,  {Couchman} H. M.~P.,  {Stinson} H. M.~P.,
  {Shen} S.,  2014, MNRAS, 442, 732

\bibitem[\protect\citeauthoryear{{Zabl} et~al.,}{{Zabl}
  et~al.}{2019}]{Zabl2019}
{Zabl} J.,  et~al., 2019, \mn@doi [\mnras] {10.1093/mnras/stz392}, \href
  {https://ui.adsabs.harvard.edu/\#abs/2019MNRAS.tmp..388Z} {p.~388}

\bibitem[\protect\citeauthoryear{{de Lapparent}, {Geller}  \& {Huchra}}{{de
  Lapparent} et~al.}{1986}]{Lapperent1986}
{de Lapparent} V.,  {Geller} M.~J.,   {Huchra} J.~P.,  1986, \mn@doi [\apjl]
  {10.1086/184625}, \href {http://adsabs.harvard.edu/abs/1986ApJ...302L...1D}
  {302, L1}

\bibitem[\protect\citeauthoryear{{van den Bergh}}{{van den
  Bergh}}{1962}]{vandenBergh1962}
{van den Bergh} S.,  1962, \mn@doi [\aj] {10.1086/108757}, \href
  {http://adsabs.harvard.edu/abs/1962AJ.....67..486V} {67, 486}

\makeatother
\end{thebibliography}





\bsp	
\label{lastpage}

 \appendix
 \section{Data}
 In Table~\ref{tbl:nofb} (\textit{no feedback} run) and Table~\ref{tbl:fb} (\textit{feedback} run) we present the  redshift evolution of the filament radius, as fitted from the density, temperature and vorticity field. 
 \newline
 \begin{table*}
 
 \begin{tabular}{|c|c|c|c|c|c|c|}
\hline
z
 & \multicolumn{3}{c|}{DM} & \multicolumn{3}{c|}{gas} \\
\cline{2-7}
 & $r_0$ (kpc) & $r_T$ (kpc) & $r_{\omega}$ (kpc)  &$r_0$ (kpc) & $r_T$ (kpc) & $r_{\omega}$ (kpc) \\
\hline
 3.35&$15.37\pm6.66$&$32.94\pm3.47$&$39.35\pm8.38$&$9.29\pm4.13$&$27.24\pm6.07$&$23.02\pm6.79$\\
 \hline
 3.44&$10.54\pm4.48$&$31.56\pm3.94$&$33.91\pm4.23$&$9.82\pm3.92$&$26.27\pm6.21$&$25.01\pm10.41$\\
 \hline
 3.60&$10.48\pm4.22$&$33.82\pm5.91$&$35.41\pm6.45$&$7.66\pm2.73$&$25.65\pm5.04$&$17.65\pm5.60$\\
 \hline
 3.65&$11.20\pm3.62$&$33.62\pm5.44$&$32.90\pm5.04$&$7.34\pm2.70$&$27.63\pm6.28$&$18.68\pm5.01$\\
 \hline
 3.82&$11.44\pm4.59$&$28.83\pm3.84$&$31.92\pm4.66$&$6.18\pm2.19$&$22.20\pm6.36$&$16.29\pm5.86$\\
 \hline
 3.94&$8.64\pm3.89$&$27.12\pm5.14$&$29.77\pm6.11$&$4.27\pm1.38$&$20.44\pm4.52$&$14.95\pm4.81$\\
 \hline
 4.00&$8.39\pm3.82$&$28.57\pm6.47$&$25.87\pm4.48$&$5.04\pm1.96$&$18.58\pm3.96$&$21.70\pm6.62$\\
 \hline
 4.13&$7.88\pm3.64$&$26.57\pm7.79$&$24.43\pm5.05$&$3.96\pm1.53$&$17.70\pm4.12$&$14.48\pm4.63$\\
 \hline
 4.55&$7.34\pm3.50$&$20.28\pm4.04$&$22.82\pm5.57$&$4.45\pm1.72$&$13.84\pm2.64$&$13.33\pm3.18$\\
 \hline
 5.06&$3.95\pm2.27$&$17.14\pm3.05$&$19.84\pm3.55$&$3.68\pm0.85$&$10.04\pm2.34$&$10.13\pm3.03$\\
 \hline
 5.45&$3.37\pm1.53$&$14.75\pm2.99$&$16.31\pm3.37$&$2.96\pm1.23$&$7.80\pm2.13$&$6.89\pm1.73$\\
 \hline
 5.66&$2.74\pm1.74$&$11.27\pm2.52$&$15.32\pm4.32$&$2.99\pm1.39$&$7.48\pm2.22$&$.23\pm1.45$\\
 \hline
 7.00&$1.75\pm0.75$&$9.19\pm2.58$&$10.33\pm1.76$&$1.77\pm0.79$&$3.79\pm0.49$&$4.46\pm0.66$\\
 \hline
 8.09&$1.59\pm1.00$&$6.28\pm1.48$&$8.18\pm1.82$&$1.58\pm0.74$&$3.78\pm0.56$&$3.96\pm0.58$\\
 \hline

\hline
\end{tabular}
\caption{Redshift evolution of the filament core radius derived from density ($r_0$) and truncation radii derived from temperature ($r_T$) and vorticity ($r_{\omega}$)  as fitted from the density, temperature and vorticity fields extracted from the \textit{no-feedback} run. See Section~\ref{sec:fit} for details. }
\label{tbl:nofb}
\end{table*}
\newline

\begin{table*}
\begin{tabular}{|c|c|c|c|c|>{\centering\arraybackslash}p{1.451cm}|>{\centering\arraybackslash}p{1.451cm}|}
\hline
z
 & \multicolumn{3}{c|}{DM} & \multicolumn{3}{c|}{gas} \\
\cline{2-7}
 & $r_0$ (kpc) & $r_T$ (kpc) & $r_{\omega}$ (kpc)  &$r_0$ (kpc) & $r_T$ (kpc) & $r_{\omega}$ (kpc) \\
\hline
 
 3.26&$10.31\pm4.41$&$36.90\pm5.36$&$35.11\pm8.25$&$7.92\pm3.77$&--&--\\
 \hline
 3.35&$10.96\pm4.81$&$33.82\pm6.43$&$33.82\pm8.34$&$12.14\pm7.02$&--&--\\
 \hline
 3.44&$10.71\pm4.77$&$34.33\pm5.42$&$34.05\pm7.36$&$8.64\pm3.95$&--&--\\
 \hline
 3.53&$16.01\pm10.50$&$34.80\pm12.75$&$38.96\pm17.90$&$7.22\pm4.23$&--&--\\
 \hline
 3.60&$11.20\pm5.17$&$36.80\pm12.47$&$35.69\pm14.22$&$8.02\pm7.38$&--&--\\
 \hline
 3.65&$11.86\pm4.88$&$30.58\pm3.67$&$31.68\pm5.53$&$8.04\pm3.75$&--&--\\
 \hline
 3.82&$10.45\pm4.37$&$28.99\pm3.93$&$30.51\pm7.28$&$5.89\pm2.79$&--&--\\
 \hline
 3.96&$10.28\pm4.59$&$27.86\pm4.01$&$24.48\pm7.15$&$5.10\pm2.29$&--&--\\
 \hline
 4.00&$7.33\pm4.43$&$27.23\pm4.84$&$22.41\pm6.05$&$5.13\pm2.28$&--&--\\
 \hline
 4.02&$9.36\pm4.30$&$26.48\pm4.34$&$24.21\pm7.01$&$4.94\pm2.34$&--&--\\
 \hline
 4.56&$8.61\pm5.60$&$23.13\pm4.57$&$23.13\pm5.06$&$4.21\pm2.10$&--&--\\
 \hline
 5.45&$3.37\pm1.96$&$14.37\pm3.83$&$12.99\pm3.13$&$2.70\pm1.55$&--&--\\
 \hline
 5.67&$2.13\pm1.33$&$11.59\pm2.57$&$11.66\pm2.60$&$2.78\pm1.50$&--&--\\
 \hline
 6.99&$1.80\pm1.13$&$9.20\pm2.69$&$9.20\pm2.58$&$2.79\pm1.51$&--&--\\
 \hline
 8.09&$1.58\pm1.00$&$7.08\pm2.19$&$6.30\pm1.91$&$2.43\pm1.49$&--&--\\
 \hline

\end{tabular}
\caption{Redshift evolution of the filament core radius derived from density ($r_0$) and truncation radii derived from temperature ($r_T$) and vorticity ($r_{\omega}$)  as fitted from the density, temperature and vorticity fields extracted from the \textit{feedback} run. See Section~\ref{sec:fit} for details.   Note the absence of data for the radii derived from vorticity and temperature, due to the destructive impact of feedback on these fields. 
}
 \label{tbl:fb}
\end{table*}

 \begin{table*}
 
 \begin{tabular}{|c|c|c|c|c|c|c|}
\hline
z
 & \multicolumn{3}{c|}{DM} & \multicolumn{3}{c|}{gas} \\
\cline{2-7}
  & $\log_{10}(\rho_0/{\rm gcm^{-3}})$ & $\log_{10}(\rho_1/{\rm g cm^{-3}})$ & $r_1 ({\rm kpc})$ & $\log_{10}(\rho_0/{\rm gcm^{-3}})$ & $\log_{10}(\rho_1/{\rm gcm^{-3}})$ & $r_1 ({\rm kpc})$\\
 \hline
 3.35&$-26.08\pm0.46$&$-26.50\pm0.46$&$7.66\pm3.15$&$-26.76\pm0.34$&$-27.32\pm0.36$&$12.36\pm5.49$\\
 \hline
 3.44&$-26.12\pm0.39$&$-26.63\pm0.40$&$9.30\pm3.62$&$-26.64\pm0.64$&$-27.19\pm1.01$&$16.32\pm7.66$\\
 \hline
 3.60&$-26.05\pm0.42$&$-26.43\pm0.44$&$7.89\pm3.20$&$-26.75\pm1.15$&$-27.26\pm0.38$&$8.69\pm3.13$\\
 \hline
 3.65&$-26.03\pm0.42$&$-26.56\pm0.43$&$8.13\pm3.00$&$-26.66\pm0.38$&$-27.30\pm0.38$&$7.82\pm2.65$\\
 \hline
 3.82&$-26.04\pm0.41$&$-26.37\pm0.48$&$8.05\pm3.00$&$-26.64\pm0.41$&$-27.25\pm0.37$&$8.82\pm3.02$\\
 \hline
 3.94&$-25.85\pm0.47$&$-26.50\pm0.41$&$7.95\pm3.42$&$-26.18\pm0.64$&$-26.95\pm1.19$&$7.68\pm2.81$\\
 \hline
 4.00&$-25.83\pm0.49$&$-26.45\pm0.45$&$6.79\pm2.80$&$-26.53\pm0.45$&$-27.12\pm0.39$&$7.70\pm3.00$\\
 \hline
 4.13&$-25.80\pm0.44$&$-26.52\pm1.10$&$6.89\pm2.75$&$-25.81\pm1.85$&$-27.05\pm0.37$&$6.76\pm2.78$\\
 \hline
 4.55&$-25.81\pm0.48$&$-26.34\pm0.41$&$6.89\pm2.97$&$-26.44\pm0.45$&$-27.00\pm0.36$&$7.17\pm2.76$\\
 \hline
 5.06&$-25.58\pm0.45$&$-26.43\pm1.20$&$6.06\pm2.80$&$-25.75\pm2.00$&$-26.88\pm0.36$&$6.27\pm2.39$\\
 \hline
 5.45&$-25.66\pm0.42$&$-26.06\pm0.61$&$6.00\pm2.55$&$-26.23\pm0.42$&$-26.66\pm1.30$&$5.92\pm2.14$\\
 \hline
 5.66&$-25.62\pm0.43$&$-26.41\pm1.25$&$9.00\pm4.78$&$-26.11\pm0.56$&$-26.92\pm0.44$&$5.47\pm2.06$\\
 \hline
 7.00&$-25.42\pm0.45$&$-26.20\pm1.20$&$6.43\pm4.50$&$-26.11\pm0.53$&$-27.09\pm0.07$&$6.39\pm4.89$\\
 \hline
 8.09&$-25.32\pm0.46$&$-25.97\pm1.22$&$6.07\pm3.69$&$-25.98\pm0.57$&$-26.66\pm0.36$&$4.53\pm1.72$\\
 \hline

\end{tabular}
\caption{Redshift evolution of the model fitted density  as extracted from the \textit{no-feedback} run. See Section~\ref{sec:fit} for details. }
\label{tbl:nofbrho}
\end{table*}

 \begin{table*}
 
 \begin{tabular}{|c|c|c|c|c|c|c|}
\hline
z
 & \multicolumn{3}{c|}{DM} & \multicolumn{3}{c|}{gas} \\
\cline{2-7}
 & $\log_{10}(\rho_0/{\rm gcm^{-3}})$ & $\log_{10}(\rho_1/{\rm g cm^{-3}})$ & $r_1 ({\rm kpc})$ & $\log_{10}(\rho_0/{\rm gcm^{-3}})$ & $\log_{10}(\rho_1/{\rm gcm^{-3}})$ & $r_1 ({\rm kpc})$\\
\hline
  3.26&$-2.26\pm0.46$&$-2.83\pm0.41$&$8.45\pm3.67$&$-26.85\pm0.42$&$-27.18\pm0.47$&$13.82\pm6.81$\\
 \hline
 3.35&$-2.40\pm0.46$&$-2.89\pm0.46$&$9.08\pm3.92$&$-26.98\pm0.50$&$-27.32\pm0.45$&$8.63\pm3.50$\\
 \hline
 3.44&$-2.36\pm0.42$&$-2.70\pm0.44$&$8.38\pm3.16$&$-26.81\pm0.45$&$-27.17\pm1.20$&$9.18\pm3.34$\\
 \hline
 3.53&$-2.49\pm0.59$&$-2.83\pm0.51$&$8.16\pm3.60$&$-26.94\pm0.69$&$-27.26\pm0.48$&$7.50\pm2.78$\\
 \hline
 3.60&$-2.38\pm0.56$&$-2.81\pm0.53$&$5.86\pm2.65$&$-26.62\pm0.86$&$-27.08\pm0.56$&$9.09\pm3.67$\\
 \hline
 3.65&$-2.31\pm0.43$&$-2.77\pm0.55$&$8.80\pm3.92$&$-26.79\pm0.41$&$-27.33\pm0.41$&$7.95\pm3.16$\\
 \hline
 3.82&$-2.26\pm0.42$&$-2.64\pm0.62$&$7.69\pm2.97$&$-26.69\pm0.39$&$-27.18\pm0.42$&$8.13\pm3.18$\\
 \hline
 3.96&$-2.23\pm0.41$&$-2.77\pm0.42$&$9.47\pm5.03$&$-26.64\pm0.42$&$-27.02\pm1.07$&$8.26\pm2.98$\\
 \hline
 4.00&$-2.19\pm0.44$&$-2.57\pm0.62$&$7.69\pm4.16$&$-26.64\pm0.43$&$-27.13\pm0.44$&$10.04\pm6.40$\\
 \hline
 4.02&$-2.21\pm0.41$&$-2.82\pm1.03$&$7.88\pm3.06$&$-26.61\pm0.43$&$-26.97\pm0.53$&$9.23\pm3.70$\\
 \hline
 4.56&$-2.04\pm0.44$&$-2.77\pm1.17$&$6.59\pm2.84$&$-26.52\pm0.41$&$-27.00\pm0.42$&$7.48\pm2.75$\\
 \hline
 5.45&$-1.82\pm0.40$&$-2.56\pm1.17$&$6.53\pm5.57$&$-26.27\pm0.49$&$-26.95\pm0.39$&$6.03\pm2.49$\\
 \hline
 5.67&$-1.84\pm0.44$&$-2.62\pm1.30$&$5.77\pm2.38$&$-26.35\pm0.50$&$-26.92\pm0.37$&$6.18\pm2.40$\\
 \hline
 6.99&$-1.59\pm0.49$&$-2.41\pm1.23$&$3.74\pm1.39$&$-26.15\pm0.46$&$-26.81\pm0.36$&$6.01\pm2.38$\\
 \hline
 8.09&$-1.49\pm0.57$&$-2.10\pm0.42$&$3.36\pm1.49$&$-26.03\pm0.57$&$-26.88\pm0.26$&$4.20\pm1.66$\\
 \hline

 \hline
\end{tabular}
\caption{Redshift evolution of the model fitted density  as extracted from the \textit{feedback} run. See Section~\ref{sec:fit} for details. }
\label{tbl:fbrho}
\end{table*}

\end{document}